\pdfoutput=1

\RequirePackage{etoolbox}
\csdef{input@path}{
 {img/}
}

\documentclass{article}

\usepackage{arxiv}

\usepackage[utf8]{inputenc} % allow utf-8 input
\usepackage[T1]{fontenc}    % use 8-bit T1 fonts
\usepackage{hyperref}       % hyperlinks
\usepackage{url}            % simple URL typesetting
\usepackage{booktabs}       % professional-quality tables
\usepackage{amsfonts}       % blackboard math symbols
\usepackage{nicefrac}       % compact symbols for 1/2, etc.
\usepackage{microtype}      % microtypography
\usepackage{lipsum}
\usepackage{amsthm}
\usepackage{amsmath}
\usepackage{amssymb}
\usepackage{titlesec}
\usepackage{graphicx}
\usepackage{mathtools}
\usepackage{gensymb}
\usepackage[nomarkers,figuresonly]{endfloat}

\title{Bayesian Inference Gaussian Process Multiproxy Alignment of Continuous Signals (BIGMACS): Applications for Paleoceanography}

\author{
  Taehee Lee \\
  Department of Statistics\\
  Harvard University\\
  Massachusetts, USA \\
  \texttt{taehee\_lee@fas.harvard.edu} \\
  \And
  Lorraine E. Lisiecki \\
  Department of Earth Science\\
  University of California, Santa Barbara\\
  California, USA \\
  \texttt{lisiecki@geol.ucsb.edu} \\
  \And
  Devin Rand \\
  Department of Earth Science\\
  University of California, Santa Barbara\\
  California, USA \\
  \texttt{drand@ucsb.edu} \\
  \And
  Geoffrey Gebbie \\
  Department of Physical Oceanography\\
  Woods Hole Oceanographic Institution\\
  Massachusetts, USA \\
  \texttt{ggebbie@whoi.edu} \\
  \And
  Charles E. Lawrence \\
  Division of Applied Mathematics\\
  Brown University\\
  Rhode Island, USA \\
  \texttt{charles\_lawrence@brown.edu} \\
}

\begin{document}
\maketitle

\begin{abstract}
We first introduce a novel profile-based alignment algorithm, the multiple continuous Signal Alignment algorithm with Gaussian Process Regression profiles (SA-GPR). SA-GPR addresses the limitations of currently available signal alignment methods by adopting a hybrid of the particle smoothing and Markov-chain Monte Carlo (MCMC) algorithms to align signals, and by applying the Gaussian process regression to construct profiles to be aligned continuously. SA-GPR shares all the strengths of the existing alignment algorithms that depend on profiles but is more exact in the sense that profiles do not need to be discretized as sequential bins. The uncertainty of performance over the resolution of such bins is thereby eliminated. This methodology produces alignments that are consistent, that regularize extreme cases, and that properly reflect the inherent uncertainty.

Then we extend SA-GPR to a specific problem in the field of paleoceanography with a method called Bayesian Inference Gaussian Process Multiproxy Alignment of Continuous Signals (BIGMACS). The goal of BIGMACS is to infer continuous ages for ocean sediment cores using two classes of age proxies: proxies that explicitly return calendar ages (e.g., radiocarbon) and those used to synchronize ages in multiple marine records (e.g., an oxygen isotope based marine proxy known as benthic ${\delta}^{18}{\rm O}$). BIGMACS integrates these two proxies by iteratively performing two steps: profile construction from benthic ${\delta}^{18}{\rm O}$ age models and alignment of each core to the profile also reflecting radiocarbon dates. We use BIGMACS to construct a new Deep Northeastern Atlantic stack (i.e., a profile from a particular benthic ${\delta}^{18}{\rm O}$ records) of five ocean sediment cores. We conclude by constructing multiproxy age models for two additional cores from the same region by aligning them to the stack.
\end{abstract}

% keywords can be removed

\section{Introduction}\label{sec1}

Paleoceanography is the study of past oceans, with particular emphasis on estimating past changes in climate. In paleoceanography, marine fossils and sediments deposited on the seafloor are recovered in ocean sediment cores and are used as a proxy for past climate states. A climate proxy is a measurement that provides an indirect means to estimate a value of interest (e.g., global ice volume). A common climate proxy in paleoceanography is the isotopic ratio of ${}^{18}{\rm O}$/${}^{16}{\rm O}$ measured in calcite fossils from the deep ocean, known as benthic ${\delta}^{18}{\rm O}$. Variations of benthic ${\delta}^{18}{\rm O}$ in an ocean sediment core indicate past changes in global ice volume, local water temperature, and local salinity. Before climate proxy signals from sediment cores can be interpreted, the sediment layer depths from which the climate proxy was measured must be converted to time estimates using an age model which relies on the principle that younger sediments are deposited on top of older sediments and that changes in sedimentation rate are gradual.

One method for constructing age models is using age estimates from dating proxies, such as radiocarbon or tephra layers (fragmental material produced by a volcanic eruption), that are observed in a subset of the sediment layers and interpolating between the dates. However, dating proxies are often limited in range and resolution. In the absence of dating proxies, the alignment of global climate proxies, such as benthic ${\delta}^{18}{\rm O}$, can be used as a dating technique. Under the assumption that ${\delta}^{18}{\rm O}$ changes synchronously in different locations, the ${\delta}^{18}{\rm O}$ signal of a core without an age model can be aligned to a ${\delta}^{18}{\rm O}$ signal that has an age model. The former core can then adopt the age model of the latter core; henceforth, we refer to climate proxies that can be used to produce alignment-based age models as synchronizing proxies. The alignment process should minimize the residuals between the synchronizing proxies while maintaining realistic sediment accumulation rates, defined as the difference between sediment layer depths divided by their age difference.

The core alignment problem in paleoceanography is related to the signal alignment problem in statistics, such as dynamic time warping (DTW) \cite{Munich1999}. Here, we have multiple cores to be aligned to one another. Pairwise alignment algorithms quickly become computationally intractable when aligning multiple cores and direct pairwise alignments may be inconsistent with the indirect alignment of the same pair with a third core. Profile analysis \cite{Durbin1998}, or motif finding \cite{Bailey2009}, circumvents these limitations by building a common probabilistic model, called a profile. The profile is assumed to generate or “emit” each of the signals. In paleoceanography, profiles of the benthic ${\delta}^{18}{\rm O}$ climate proxy are called stacks (e.g., \cite{Lisiecki2005,Ahn2016}). Each core is aligned to a given stack for assigning ages to its benthic ${\delta}^{18}{\rm O}$ observations. Ideally, the stack should be updated based on all available alignments, and then the stacking and alignment steps should be iterated until convergence. Ultimately, the constructed stack serves a probabilistic summary of benthic ${\delta}^{18}{\rm O}$ from the aligned cores and a target for future benthic ${\delta}^{18}{\rm O}$ alignments.

Dating proxies and synchronizing proxies are complementary in the dating process. Dating proxies causally relate to age but generally have lower resolution, whereas synchronizing proxies often have higher resolution but can only provide age information when compared across cores. Thus, a comprehensive algorithm that utilizes both types of proxies simultaneously has the potential to improve existing dating techniques. There have been many alignment algorithms in computational biology but they deal with bases that are discrete whereas ages and climate variations are continuous by nature, meaning that ages are drawn from a continuous domain such as real line intervals rather than from a finite discrete space such as 4 nucleotides or 20 amino acids.

The profile hidden Markov model (HMM) is a popular profile analysis for aligning multiple signals or sequences in computational biology \cite{Durbin1998,Eddy1998,Ewens2013} or human speech recognition \cite{Baker1975,Huang2001,Gales2008}. It iterates to align each signal to the profile by the forward-backward algorithm and to update the profile based on the signal alignments by the EM algorithm. Because it assumes that alignments are drawn from a finite set of candidates over which the profile is defined and the emission model at each position in the profile is updated directly from the samples that are assigned to that position, the profile HMM is limited to dealing with continuous alignments and uneven data resolution, such as dating proxies in paleoceanography.

The DTW algorithm \cite{Munich1999,Hay2019,Hagen2020} is conceptually identical to the Needleman-Wunsch algorithm \cite{Needleman1970} in bioinformatics; it searches the optimal warping paths for the pair of query signals by solving the associated dynamic programming or forward-backward algorithm and returns the associated score. Though some of its variations, such as DTW barycenter averaging (DBA) \cite{Petitjean2011,Petitjean2014} methods, use the centroid signal as a profile to be aligned to overcome the limitation from the pairwise alignment aspect of DTW algorithms, it is still limited in some problems including ours; DTW algorithms are originally designed for discrete data with integer positions to be exactly aligned one-by-one but proxy measurements sampled from core sediments are intrinsically asynchronous. Also, the performance of DTW is sensitive to the penalties and loss functions in the transition and emission models. The stochastic alignment process (SAP) \cite{Juang1984,Nakagawa1988,Nakagawa1989,Chudova2003,Marteau2019} is a hybrid approach of the HMM and DTW that adopts probabilistic transition and emission models and aims to either get the most likely warping path or sample warping paths given data, but it is still designed to use discrete time and states and, therefore, not well suited for continuous alignments, as it shares the limitations of HMM and DTW.

In this paper, we introduce a new framework of the multiple signal alignment algorithm, the multiple continuous Signal Alignment algorithm with Gaussian Process Regression profiles (SA-GPR), and its practical application to paleoceanography, Bayesian Inference Gaussian Process Multiproxy Alignment of Continuous Signals (BIGMACS). BIGMACS is designed to infer ages from radiocarbon (an dating proxy) and benthic ${\delta}^{18}{\rm O}$ (a synchronizing proxy) simultaneously by incorporating calendar ages into the alignment process. A hybrid of the particle smoothing \cite{Doucet2001} and MCMC algorithms \cite{Martino2015} probabilistically aligns signals (benthic ${\delta}^{18}{\rm O}$) to the given profile (stack) by sampling the alignments continuously. The profile (stack) is continuously updated as the integration of Gaussian process regression \cite{Rasmussen2005} over the sampled alignments.

Section \ref{sec2} covers the structure, models and inference of SA-GPR. Section \ref{sec3} defines the application to paleoceanography, lists existing methods in the field with their limitations, and describes terminologies of paleoceanography and statistical models that we adopt for radiocarbon and benthic ${\delta}^{18}{\rm O}$ proxies. Section \ref{sec4} shows some concrete application results of the core alignments and stack construction. The discussion and conclusion are given in sections \ref{sec5} and \ref{sec6}, respectively. All detailed computations and supporting materials are found in the supplementary notes.

\section{SA-GPR: Multiple Continuous Signal Alignment Algorithm}\label{sec2}

The multiple continuous signal alignment algorithm with Gaussian process regression profiles (SA-GPR) is best described as a framework rather than a software. This Bayesian framework allows the sampling of alignments and construction of a profile that integrates the sampled alignments, rather than depending on point estimates. All detailed formulations are covered in Section S2 of the supplementary notes.

\subsection{Models}\label{sec2-1}

SA-GPR consists of two steps: the signal alignment step and profile construction step. The signal alignment step samples alignments to the given profile and profile construction step updates the profile based on the given alignment samples. These two steps are iterated until convergence.

Suppose that we have $\rm{M}$ signals to align and each signal $m$ consists of ${\rm{N}}_{m}$ pairs of inputs and outputs, or positions and observations more straightforwardly, ${\rm{X}}^{\left(m\right)}=\left\{x_n^{\left(m\right)}\right\}_{n=1}^{{\rm{N}}_m}$ and ${\rm{Y}}^{\left(m\right)}=\left\{y_n^{\left(m\right)}\right\}_{n=1}^{{\rm{N}}_m}$, respectively, where each $x_n^{\left(m\right)}\in\mathcal{X}$ and $y_n^{\left(m\right)}\in\mathcal{Y}$ for some (measurable) spaces $\mathcal{X}$ and $\mathcal{Y}$.

\subsubsection{Signal Alignment}\label{sec2-1-1}

The signal alignment step aims to infer the alignment ${\rm{Z}}^{\left(m\right)}=\left\{{\rm{Z}}_n^{\left(m\right)}\right\}_{n=1}^{{\rm{N}}_m}$ for each signal $m$ to the given profile $\mathcal{P}=\left(g,\varphi\right)$, where each ${\rm{Z}}_n^{\left(m\right)}$ is defined on a single ordered field $\mathcal{S}$ and $g_\varphi$ is a probability density function on $\mathcal{Y}$ that is depending on the function $\varphi$ over $\mathcal{S}$. The modelling can be understood as a state-space model.

First, we have a prior on the hidden variable ${\rm{Z}}^{\left(m\right)}$ given ${\rm{X}}^{\left(m\right)}$ as follows:
\begin{equation}
p\left({\rm{Z}}^{\left(m\right)}\middle|{\rm{X}}^{\left(m\right)},\theta^{\left(m\right)}\right)=\pi_0\left({\rm{Z}}_1^{\left(m\right)}\middle| x_1^{\left(m\right)},\theta^{\left(m\right)}\right)\prod_{n=2}^{{\rm{N}}_m}{\pi\left({\rm{Z}}_n^{\left(m\right)}\middle|{\rm{Z}}_{n-1}^{\left(m\right)},x_n^{\left(m\right)},x_{n-1}^{\left(m\right)},\theta^{\left(m\right)}\right)},
\label{equ2-1-1}
\end{equation}
\begin{equation}
\theta^{\left(m\right)} \sim \pi_{tr},
\label{equ2-1-2}
\end{equation}
where $\pi_0$ and $\pi$ form a transition model in the terminology of the state-space modelling. These components depend on the parameter $\theta^{\left(m\right)}$ that follows the prior $\pi_{tr}$. The term $\pi\left({\rm{Z}}_n^{\left(m\right)}\middle|Z_{n-1}^{\left(m\right)},x_n^{\left(m\right)},x_{n-1}^{\left(m\right)}\right)$ implies that the distribution of the nth alignment ${\rm{Z}}_n^{\left(m\right)}$ given ${\rm{Z}}_{n-1}^{\left(m\right)}$ and ${\rm{X}}^{\left(m\right)}$ is regularized by not only the previous alignment ${\rm{Z}}_{n-1}^{\left(m\right)}$ but also the corresponding positions $x_n^{\left(m\right)}$ and $x_{n-1}^{\left(m\right)}$. For example, the distance between ${\rm{Z}}_n^{\left(m\right)}$ and ${\rm{Z}}_{n-1}^{\left(m\right)}$ could be correlated with the distance between $x_n^{\left(m\right)}$ and $x_{n-1}^{\left(m\right)}$.

Second, the likelihood of ${\rm{Y}}^{\left(m\right)}$ given ${\rm{Z}}^{\left(m\right)}$ is as follows:
\begin{equation}
p\left({\rm{Y}}^{\left(m\right)}\middle|{\rm{Z}}^{\left(m\right)},\psi^{\left(m\right)};\mathcal{P}\right)=\prod_{n=1}^{{\rm{N}}_m}{g\left(y_n^{\left(m\right)}\middle|\varphi\left({\rm{Z}}_n^{\left(m\right)},\psi^{\left(m\right)}\right)\right)},
\label{equ2-1-3}
\end{equation}
\begin{equation}
\psi^{\left(m\right)} \sim \pi_{em},
\label{equ2-1-4}
\end{equation}
i.e., ${\rm{Y}}^{\left(m\right)}$ is conditionally independent from the other variables and other sequences given the alignment ${\rm{Z}}^{\left(m\right)}$ and the profile $\mathcal{P}$ that defines the sufficient statistic $\varphi$, and its modelling is depending on the parameter $\psi^{\left(m\right)}$ that has the prior $\pi_{em}$. Equation (\ref{equ2-1-3}) is often called an emission model in the terminology of the state-space modelling.

The goal of the signal alignment step is to sample the alignment ${\rm{Z}}^{\left(m\right)}$ from its posterior after estimating ${\hat{\theta}}^{\left(m\right)}=\theta^{\left(m\right)}$ and ${\hat{\psi}}^{\left(m\right)}=\psi^{\left(m\right)}$:
\begin{equation}
\begin{aligned}
{\rm{Z}}^{\left(m\right)} & \sim p\left({\rm{Z}}^{\left(m\right)}\middle|{\rm{X}}^{\left(m\right)},{\rm{Y}}^{\left(m\right)},{\hat{\theta}}^{\left(m\right)}, {\hat{\psi}}^{\left(m\right)}\right) \\ & \propto p\left({\rm{Z}}^{\left(m\right)}\middle|{\rm{X}}^{\left(m\right)}, {\hat{\theta}}^{\left(m\right)}\right)p\left({\rm{Y}}^{\left(m\right)}\middle|{\rm{Z}}^{\left(m\right)}, {\hat{\psi}}^{\left(m\right)};\mathcal{P}\right).
\end{aligned}
\label{equ2-1-5}
\end{equation}

Sampling of ${\rm{Z}}^{\left(m\right)}$ and estimation of $\theta^{\left(m\right)}$ and  $\psi^{\left(m\right)}$ will be discussed in section \ref{sec2-2}.

\subsubsection{Profile Construction}\label{sec2-1-2}

Suppose that a set of $\rm{L}$ samples $\widetilde{Z}=\left\{{\widetilde{\rm{Z}}}_l\right\}_{l=1}^{\rm{L}}$, where ${\widetilde{\rm{Z}}}_l=\left\{{\widetilde{\rm{Z}}}^{\left(m,l\right)}\right\}_{m=1}^{\rm{M}}$ and ${\widetilde{Z}}^{\left(m,l\right)}$ was drawn from the posterior in (\ref{equ2-1-5}) for each signal m. For each pair $\left({\rm{Z}},{\rm{Y}}\right)$, let us assume that a probabilistic model $p\left(y\middle| z;{\rm{Z}},{\rm{Y}}\right)$ is defined for any pair of $\left(z,y\right)\in\mathcal{S}\times\mathcal{Y}$. Then, the profile is defined by the posterior predictive, which is obtained by marginalizing out the hidden variable $\rm{Z}$:
\begin{equation}
\begin{aligned}
p\left(y\middle| z;{\rm{Y}},{\rm{X}}\right) & =\int{p\left(y\middle| z;{\rm{Z}},{\rm{Y}}\right)p\left({\rm{Z}}\middle|{\rm{X}},{\rm{Y}},\hat{\theta},\hat{\psi}\right)d{\rm{Z}}} \\ & =\int{p\left(y\middle| z;{\rm{Z}},{\rm{Y}}\right)\prod_{m=1}^{\rm{M}}{p\left({\rm{Z}}^{\left(m\right)}\middle|{\rm{X}}^{\left(m\right)},{\rm{Y}}^{\left(m\right)},{\hat{\theta}}^{\left(m\right)},\ {\hat{\psi}}^{\left(m\right)}\right)}d{\rm{Z}}} \\ & \approx \frac{1}{\rm{L}}\sum_{l=1}^{\rm{L}}{p\left(y\middle| z;{\widetilde{\rm{Z}}}_l,{\rm{Y}}\right)}.
\end{aligned}
\label{equ2-1-6}
\end{equation}

Any model can be plugged in (\ref{equ2-1-6}). Here, for the case $\mathcal{Y}=\mathbb{R}$ and a metric space $\mathcal{S}$, we consider the Gaussian process regression (GPR) \cite{Rasmussen2005}, which is a model for nonparametric regression. A GPR is defined by its mean prior function $\underline{\mu}:\mathcal{S}\rightarrow\mathbb{R}$ and kernel function $\mathbb{K}:\mathcal{S}\times\mathcal{S}\rightarrow\mathbb{R}$. Because the exact GPR quickly becomes intractable once the size of training data grows since its time complexity is $\mathcal{O}\left(\left|{\rm{Z}}\right|^3\right)$, it is inevitable to resort to the variational approximations \cite{Bauer2016}. The variational free energy (VFE) \cite{Titsias2009} gives the following explicit form of the model $p\left(y\middle| z;{\rm{Z}},{\rm{Y}}\right)$:
\begin{equation}
p\left(y\middle| z;{\rm{Z}},{\rm{Y}}\right)=\mathcal{N}\left(y\middle|\overline{\mu}\left(z\right),\overline{\nu}\left(z\right)\right),
\label{equ2-1-7}
\end{equation}
where:
\begin{equation}
\begin{aligned}
\overline{\mu}\left(z\right) & = {\underline{\mu}}_z+\mathbb{K}_{z\underline{\rm{Z}}}\left(\mathbb{K}_{\underline{\rm{Z}}\underline{\rm{Z}}}+\mathbb{K}_{\underline{\rm{Z}}{\rm{Z}}}\Lambda_{\rm{Z}}^{-1}\mathbb{K}_{{\rm{Z}}\underline{\rm{Z}}}\right)^{-1}\mathbb{K}_{\underline{\rm{Z}}{\rm{Z}}}\Lambda_{\rm{Z}}^{-1}\left({\rm{Y}}-{\underline{\mu}}_{\rm{Z}}\right), \\ \overline{\nu}\left(z\right) & = \mathbb{K}_{zz}+\Lambda\left(z\right)-\mathbb{K}_{z\underline{\rm{Z}}}\mathbb{K}_{\underline{\rm{Z}}\underline{\rm{Z}}}^{-1}\mathbb{K}_{\underline{\rm{Z}}z}+\mathbb{K}_{z\underline{\rm{Z}}}\left(\mathbb{K}_{\underline{\rm{Z}}\underline{\rm{Z}}}+\mathbb{K}_{\underline{\rm{Z}}{\rm{Z}}}\Lambda_{\rm{Z}}^{-1}\mathbb{K}_{{\rm{Z}}\underline{\rm{Z}}}\right)^{-1}\mathbb{K}_{\underline{\rm{Z}}z},
\end{aligned}
\label{equ2-1-8}
\end{equation}
where ${\underline{\mu}}_{\rm{Z}}$ is a vector form of the function values of $\underline{\mu}$ at ${\rm{Z}}$, $\mathbb{K}_{\rm{ZZ}}$ is a matrix form of the function values of $\mathbb{K}$ at ${\rm{Z}}\times {\rm{Z}}$, $\Lambda_{\rm{Z}}$ is a diagonal matrix of which diagonal entries are the values of the observational variance function $\Lambda$ at ${\rm{Z}}$, and $\underline{\rm{Z}}\subseteq\mathcal{S}$ is a pseudo-input, which is a small subset of 
$\mathcal{S}$. The time complexity of (\ref{equ2-1-8}) is $\mathcal{O}\left(\left|\underline{\rm{Z}}\right|^2\left|{\rm{Z}}\right|\right)$. It is possible to get the regression like (\ref{equ2-1-7}) and (\ref{equ2-1-8}) if $\mathcal{Y}=\mathbb{R}^{\rm{D}}$ by the multivariate Gaussian process regression \cite{chen2020}, but it is beyond the scope of this paper.

To define the stack succinctly, (\ref{equ2-1-6}) with (\ref{equ2-1-7}) is reduced to the following Gaussian distribution based on the moment-matching \cite{Murphy2012}:
\begin{equation}
\begin{aligned}
& p\left(y\middle| z;{\rm{Y}},{\rm{X}}\right)=\mathcal{N}\left(y\middle|\overline{\mu}\left(z\right),\overline{\nu}\left(z\right)\right) \\ & \approx\mathcal{N}\left(y\middle|\frac{1}{\rm{L}}\sum_{l=1}^{\rm{L}}{{\overline{\mu}}^{\left(l\right)}\left(z\right)},\frac{1}{\rm{L}}\sum_{l=1}^{\rm{L}}\left({\overline{\nu}}^{\left(l\right)}\left(z\right)+\Lambda^{\left(l\right)}\left(z\right)+\left({\overline{\mu}}^{\left(l\right)}\left(z\right)-\frac{1}{\rm{L}}\sum_{l=1}^{\rm{L}}{{\overline{\mu}}^{\left(l\right)}\left(z\right)}\right)^2\right)\right),
\end{aligned}
\label{equ2-1-9}
\end{equation}
where ${\overline{\mu}}^{\left(l\right)}$, ${\overline{\nu}}^{\left(l\right)}$ and $\Lambda^{\left(l\right)}$ are from $p\left(y\middle| z;{\widetilde{\rm{Z}}}_l,{\rm{Y}}\right)=\mathcal{N}\left(y\middle|{\overline{\mu}}^{\left(l\right)}\left(z\right),{\overline{\nu}}^{\left(l\right)}\left(z\right)\right)$ in (\ref{equ2-1-7}) and (\ref{equ2-1-8}).

\subsection{Inference Algorithm}\label{sec2-2}

To estimate $\theta^{\left(m\right)}$ and $\psi^{\left(m\right)}$ for each signal $m$, we consider the expectation-maximization (EM) algorithm \cite{Dempster1977}, by iterating the following steps:

\begin{itemize}
  \item E-step: define the following ${\rm{Q}}\left(\theta^{\left(m\right)},\psi^{\left(m\right)}\middle|\theta^{\left(m,t\right)},\psi^{\left(m,t\right)}\right)$;
  \begin{equation}
  \begin{aligned}
& {\rm{Q}}\left(\theta^{\left(m\right)},\psi^{\left(m\right)}\middle|\theta^{\left(m,t\right)},\psi^{\left(m,t\right)}\right) \\ & =\mathbb{E}_{{\rm{Z}}^{\left(m\right)}\left|{\rm{X}}^{\left(m\right)},{\rm{Y}}^{\left(m\right)},\theta^{\left(m,t\right)},\psi^{\left(m,t\right)}\right.}\left[\log{p\left({\rm{Z}}^{\left(m\right)},{\rm{Y}}^{\left(m\right)}, \theta^{\left(m\right)},\psi^{\left(m\right)}\middle|{\rm{X}}^{\left(m\right)},\mathcal{P}\right)}\right].
\end{aligned}
\label{equ2-2-1}
\end{equation}
  \item M-step: find $\left(\theta^{\left(m,t+1\right)},\psi^{\left(m,t+1\right)}\right)$ that maximizes ${\rm{Q}}\left(\theta^{\left(m\right)},\psi^{\left(m\right)}\middle|\theta^{\left(m,t\right)},\psi^{\left(m,t\right)}\right)$;
  \begin{equation}
\left(\theta^{\left(m,t+1\right)},\psi^{\left(m,t+1\right)}\right)=  \underset{\left(\theta^{\left(m\right)},\psi^{\left(m\right)}\right)}{\arg\max}{{\rm{Q}}\left(\theta^{\left(m\right)},\psi^{\left(m\right)}\middle|\theta^{\left(m,t\right)},\psi^{\left(m,t\right)}\right)}.
\label{equ2-2-2}
\end{equation}
\end{itemize}

To compute the above ${\rm{Q}}\left(\theta^{\left(m\right)},\psi^{\left(m\right)}\middle|\theta^{\left(m,t\right)},\psi^{\left(m,t\right)}\right)$, we approximate it from the samples $\left\{{\widetilde{Z}}^{\left(m,l,t\right)}\right\}_{l=1}^{\rm{L}}$ drawn from the posterior $p\left({\rm{Z}}^{\left(m\right)}\middle|{\rm{X}}^{\left(m\right)},{\rm{Y}}^{\left(m\right)},\theta^{\left(m,t\right)},\psi^{\left(m,t\right)},\mathcal{P}\right)$ independently:
\begin{equation}
\begin{aligned}
& {\rm{Q}}\left(\theta^{\left(m\right)},\psi^{\left(m\right)}\middle|\theta^{\left(m,t\right)},\psi^{\left(m,t\right)}\right) \approx\frac{1}{\rm{L}}\sum_{l=1}^{\rm{L}}\log{p\left({\widetilde{\rm{Z}}}^{\left(m,l,t\right)},{\rm{Y}}^{\left(m\right)}, \theta^{\left(m\right)},\psi^{\left(m\right)}\middle|{\rm{X}}^{\left(m\right)},\mathcal{P}\right)} \\ & =\log{\pi_{tr}\left(\theta^{\left(m\right)}\right)}+\log{\pi_{em}\left(\psi^{\left(m\right)}\right)} \\ & +\frac{1}{\rm{L}}\sum_{l=1}^{\rm{L}}\left(\log{p\left({\rm{Y}}^{\left(m\right)}\middle|{\widetilde{\rm{Z}}}^{\left(m,l,t\right)},\psi^{\left(m\right)};\mathcal{P}\right)}+\log{p\left({\widetilde{\rm{Z}}}^{\left(m,l,t\right)}\middle|{\rm{X}}^{\left(m\right)},\theta^{\left(m\right)}\right)}\right).
\end{aligned}
\label{equ2-2-3}
\end{equation}

To sample ${\rm{Z}}^{\left(m\right)}$ from its posterior $p\left({\rm{Z}}^{\left(m\right)}\middle|{\rm{X}}^{\left(m\right)},{\rm{Y}}^{\left(m\right)},\theta^{\left(m\right)}, \psi^{\left(m\right)}\right)$, we first initialize it by the particle smoothing algorithm \cite{Doucet2001,Klaas2006}, which is a variational algorithm based on sequential importance sampling, by considering that $p\left({\rm{Z}}^{\left(m\right)}\middle|{\rm{X}}^{\left(m\right)},\theta^{\left(m\right)}\right)$ and $p\left({\rm{Y}}^{\left(m\right)}\middle|{\rm{Z}}^{\left(m\right)},\psi^{\left(m\right)};\mathcal{P}\right)$ are the transition and emission models, respectively, of the state-space model.

Though the particle smoothing algorithm is designed to sample from the posterior continuously, its variational aspects do not guarantee that the sampled alignments are continuous. In practice, only a few particles could appear for most of the sample alignments. To overcome this drawback, we consider Metropolis-Hastings algorithm \cite{Metropolis1953,Hastings1970,Martino2015}, which is one of the MCMC \cite{Peters2008} methods to “refine” the initialized samples given by the particle smoothing. Detailed formulations of the particle smoothing algorithm and Metropolis-Hastings algorithm are in the supplementary notes.

To sum up, the signal alignment step consists of the following two sub-steps:

Step 1. Sample $\left\{{\widetilde{\rm{Z}}}^{\left(m,l,t\right)}\right\}_{l=1}^{\rm{L}}$ from $p\left({\rm{Z}}^{\left(m\right)}\middle|{\rm{X}}^{\left(m\right)},{\rm{Y}}^{\left(m\right)},\theta^{\left(m,t\right)},\psi^{\left(m,t\right)},\mathcal{P}\right)$ by the hybrid of the particle smoothing algorithm and Metropolis-Hastings algorithm.

Step 2. Update $\theta^{\left(m,t+1\right)},\psi^{\left(m,t+1\right)}$ by the maximizer of ${\rm{Q}}\left(\theta^{\left(m\right)},\psi^{\left(m\right)}\middle|\theta^{\left(m,t\right)},\psi^{\left(m,t\right)}\right)$ approximated in (\ref{equ2-2-3}).

For the profile construction step, tuning kernel hyperparameters of $\mathbb{K}$ and the observational variance function $\Lambda$ greatly influence the performance of GPR. \cite{Titsias2009} recommends the following objective function to maximize for the kernel hyperparameter tuning: $\mathcal{L}$ is defined by each sample ${\widetilde{\rm{Z}}}_l$ that replaces $\rm{Z}$ in (\ref{equ2-2-4}).
\begin{equation}
\mathcal{L}=\log{\mathcal{N}\left({\rm{Y}}\middle|{\underline{\mu}}_{\rm{Z}},\Lambda_{\rm{Z}}+\mathbb{K}_{{\rm{Z}}\underline{\rm{Z}}}\mathbb{K}_{\underline{\rm{Z}}\underline{\rm{Z}}}^{-1}\mathbb{K}_{\underline{\rm{Z}}{\rm{Z}}}\right)}-\frac{1}{2} \cdot  trace\left(\Lambda_{\rm{Z}}^{-1}\left(\mathbb{K}_{\underline{\rm{Z}}\underline{\rm{Z}}}-\mathbb{K}_{{\rm{Z}}\underline{\rm{Z}}}\mathbb{K}_{\underline{\rm{Z}}\underline{\rm{Z}}}^{-1}\mathbb{K}_{\underline{\rm{Z}}{\rm{Z}}}\right)\right).
\label{equ2-2-4}
\end{equation}

If the variance function $\Lambda$ is constant, then the resulting GPR is called a homoscedastic GPR and $\Lambda$ can be tuned by maximizing $\mathcal{L}$ by considering it as another tuning parameter. If $\Lambda$ is not a constant function, then the resulting GPR is called a heteroscedastic GPR. In this case tuning is not straightforward unless $\Lambda$ is given a priori. Various publications have tackled the heteroscedastic case; for more details, see \cite{le2005,Kersting2007,Lazaro-Gredilla2011,Lee2019}.

The GPR is a nonparametric method and thus susceptible to outliers. In the middle of the stack construction, we classify and discard outliers regularly. Let ${\rm{O}}^{\left(m\right)}=\left\{{\rm{O}}_n^{\left(m\right)}\right\}_{n=1}^{{\rm{N}}_m}$ be a set of latent variables that indicate outliers, where ${\rm{O}}_n^{\left(m\right)}=1$ if the associated position-observation pair $\left(x_n^{\left(m\right)},y_n^{\left(m\right)}\right)$ is considered as an outlier, 0 otherwise.

Let us rigorously define outliers as data that do not follow the main Gaussian emission model $p\left(y\middle| z;{\rm{Y}},{\rm{X}}\right)=\mathcal{N}\left(y\middle|\overline{\mu}\left(z\right),\overline{\nu}\left(z\right)\right)$ in (\ref{equ2-1-9}), but instead follow an alternative model $\widetilde{g}$. Also assume that outliers are independent from the inputs given ${\rm{Z}}$, i.e., we have the following prior (\ref{equ2-2-5}) and likelihood (\ref{equ2-2-6}) for a small positive hyperparameter $\delta>0$:
\begin{equation}
{\rm{O}}_n^{\left(m\right)} {\sim}_{i.i.d.} Bernoulli\left(\delta\right),
\label{equ2-2-5}
\end{equation}
\begin{equation}
p\left(y_n^{\left(m\right)}\middle|{\rm{Z}}_n^{\left(m\right)},{\rm{O}}_n^{\left(m\right)}\right)=\begin{cases}
        p\left(y_n^{\left(m\right)}\middle|{\rm{Z}}_n^{\left(m\right)};{\rm{Y}},{\rm{X}}\right), & {\rm{O}}_n^{\left(m\right)}=0 \\ \widetilde{g}\left(y_n^{\left(m\right)}\middle|{\rm{Z}}_n^{\left(m\right)}\right), & {\rm{O}}_n^{\left(m\right)}=1 \end{cases}.
\label{equ2-2-6}
\end{equation}

Then, one can easily get the posterior of ${\rm{O}}_n^{\left(m\right)}$ from (\ref{equ2-2-5}) and (\ref{equ2-2-6}) as follows:
\begin{equation}
p\left({\rm{O}}_n^{\left(m\right)}=1\middle|{\rm{Z}}_n^{\left(m\right)},y_n^{\left(m\right)}\right)=\frac{\delta\cdot\widetilde{g}\left(y_n^{\left(m\right)}\middle|{\rm{Z}}_n^{\left(m\right)}\right)}{\delta\cdot\widetilde{g}\left(y_n^{\left(m\right)}\middle|{\rm{Z}}_n^{\left(m\right)}\right)+\left(1-\delta\right)p\left(y_n^{\left(m\right)}\middle|{\rm{Z}}_n^{\left(m\right)};{\rm{Y}},{\rm{X}}\right)}.
\label{equ2-2-7}
\end{equation}

Here we define the alternative model $\widetilde{g}$ as follows:
\begin{equation}
\widetilde{g}\left(y\middle| z\right)=\frac{1}{2}\cdot\mathcal{N}\left(y\middle|\overline{\mu}\left(z\right)+3\sqrt{\overline{\nu}\left(z\right)},\overline{\nu}\left(z\right)\right)+\frac{1}{2}\cdot\mathcal{N}\left(y\middle|\overline{\mu}\left(z\right)-3\sqrt{\overline{\nu}\left(z\right)},\overline{\nu}\left(z\right)\right).
\label{equ2-2-8}
\end{equation}

Thus, one can sample or infer ${\rm{O}}_n^{\left(m\right)}$ for each sampled ${\rm{Z}}_n^{\left(m\right)}$.

To sum up, the profile construction step consists of the following two sub-steps:

Step 1. Sample ${\rm{O}}^{\left(m,l\right)}$ from (\ref{equ2-2-7}) for each signal $m$ and randomly draw ${\underline{\rm{Z}}}_l$ from $\mathcal{S}$ for each sample $l$.

Step 2. Tune kernel hyperparameters of $\mathbb{K}$ and observational variance function $\Lambda^{\left(l\right)}$ for each sample $l$.

Step 3. Compute the GPR $p\left(y\middle| z;{\widetilde{\rm{Z}}}_l,{\rm{Y}}\right)=\mathcal{N}\left(y\middle|{\overline{\mu}}^{\left(l\right)}\left(z\right),{\overline{\nu}}^{\left(l\right)}\left(z\right)\right)$ defined in (\ref{equ2-1-7}) and (\ref{equ2-1-8}) for each sample $l$.

Step 4. Update the profile $p\left(y\middle| z;{\rm{Y}},{\rm{X}}\right)=\mathcal{N}\left(y\middle|\overline{\mu}\left(z\right),\overline{\nu}\left(z\right)\right)$ defined in (\ref{equ2-1-9}).

\subsection{Simulation Results}\label{sec2-3}

In all of the following toy examples, the profile is defined on $\mathcal{S}=\left[-1,1\right]$ and initialized with the first query signal. We adopt the following Gamma distribution for defining the prior (transition model) on the hidden alignment variables:
\begin{equation}
\begin{aligned}
\pi & \left({\rm{Z}}_n^{\left(m\right)}\middle|{\rm{Z}}_{n-1}^{\left(m\right)},x_n^{\left(m\right)},x_{n-1}^{\left(m\right)},\alpha^{\left(m\right)},\beta^{\left(m\right)},r^{\left(m\right)}\right) \\ & \propto {\rm{Gamma}}\left(\frac{{\rm{Z}}_n^{\left(m\right)}-{\rm{Z}}_{n-1}^{\left(m\right)}}{r^{\left(m\right)}\left(x_n^{\left(m\right)}-x_{n-1}^{\left(m\right)}\right)}\middle|\alpha^{\left(m\right)},\beta^{\left(m\right)}\right),
\end{aligned}
\label{equ2-3-1}
\end{equation}
\begin{equation}
\log{\pi_{tr}\left(\alpha^{\left(m\right)},\beta^{\left(m\right)}\right)}\propto\left(\alpha^{\left(m\right)}-1\right)\log{\underline{p}}-\underline{r}\log{\Gamma\left(\alpha^{\left(m\right)}\right)}+\alpha^{\left(m\right)}\underline{s}\log{\beta^{\left(m\right)}}-\beta^{\left(m\right)}\underline{q},
\label{equ2-3-2}
\end{equation}
where $\underline{p}=1$, $\underline{q}=5$, $\underline{r}=5$ and $\underline{s}=5$ are fixed hyperparameters \cite{Miller1980,fink1997}. $r^{\left(m\right)}$ is a depth-scale parameter that rescales ${\rm{X}}^{\left(m\right)}$.

To make the sampling robust to the outliers, we redefine the emission model that is given to be Gaussian by the following generalized Student’s t-distribution \cite{Christen2009} with a signal-specific shift parameter $h^{\left(m\right)}$:
\begin{equation}
g\left(y_n^{\left(m\right)}\middle|\varphi\left({\rm{Z}}_n^{\left(m\right)},h^{\left(m\right)}\right)\right)=\mathcal{T}_{2a}\left(y_n^{\left(m\right)}\middle|\overline{\mu}\left({\rm{Z}}_n^{\left(m\right)}\right)+h^{\left(m\right)},\sqrt{\frac{b}{a} \cdot  \overline{\nu}\left({\rm{Z}}_n^{\left(m\right)}\right)}\right),
\label{equ2-3-3}
\end{equation}
\begin{equation}
\log{\pi_{em}\left(h^{\left(m\right)}\right)}\propto-\frac{1}{2}\left(\frac{h^{\left(m\right)}-\underline{h}}{\underline{\sigma}}\right)^2,
\label{equ2-3-4}
\end{equation}
where $a=3$, $b=4$, $\underline{h}=0$, $\underline{\sigma}=1$ are the fixed hyperparameters. Here, $\mathcal{T}_{2a}\left(x\middle|\mu,\sigma\right)$ in (\ref{equ2-3-3}) is the value of a probability density function at $x$ where $\mu$ and $\sigma$ as the location and scale parameters and $2a$ is the degree of freedom.

In profile construction, we set $\delta=0.05$ in (\ref{equ2-2-5}) for classifying and discarding outliers, and pseudo-inputs $\underline{\rm{Z}}=\left\{{\underline{\rm{Z}}}_l\right\}_{l=1}^{\rm{L}}$ are chosen such that each ${\underline{\rm{Z}}}_l$ is a randomly chosen subset of $\mathcal{S}=\left[-1,1\right]$. We assume that kernel hyperparameters are shared over the samples, but the observational variance, $\Lambda^{\left(l\right)}$, is tuned sample-specifically for each $l$. For the heteroscedastic GPR, we adopt the method in \cite{Lee2019}. The Ornstein-Uhlenbeck (OU) kernel \cite{Rasmussen2005} was chosen as the kernel function. MATLAB codes that reproduce the results are available in the following link: \url{http://github.com/eilion/SA-GPR}.

\subsubsection{Example 1: Asynchronous Query Signals}\label{sec2-3-1}

One of the major defects of alignment algorithms based on the dynamic time warping (DTW) is that the results are highly dependent on the choice of input-output pairs in each query signal. If such pairs are not synchronous across different signals, the results would not be proper even if each signal is sampled from the same sequence.

\begin{figure}
\centering
\includegraphics[width=1\textwidth]{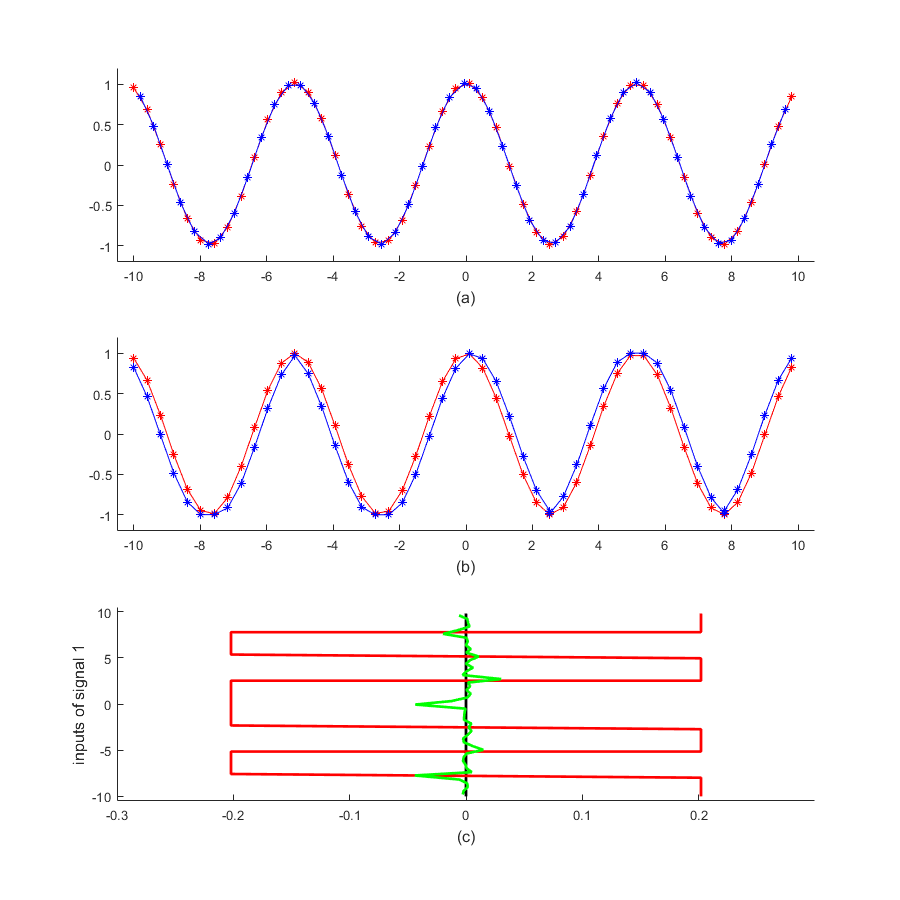}
\caption{(a) Median alignments obtained by SA-GPR. (b) Alignments obtained by DTW. (c) Errors of inferred alignments to the true ones. Red and green graphs are those of DTW and SA-GPR, respectively.}
\label{fig2-2}
\end{figure}

Figure \ref{fig2-2} and S3 (in the supplementary notes) show how asynchronous query signals can be aligned by the SA-GPR compared to DTW. We first pick two asynchronous sets of pairs of inputs and outputs from the same cosine curve, one by one, at intervals of $0.2$. Therefore, two signals are assumed to be aligned completely. Figure \ref{fig2-2} shows the alignment results obtained by the SA-GPR (a) and DTW (b). In Figure \ref{fig2-2}, a SA-GPR profile is constructed from two signals and each is aligned probabilistically, i.e., in the form of samples, and medians of those samples are plotted. Because it is a profile-dependent alignment algorithm, the resulting alignments are consistent to the assumption that two signals are identical, whether or not the signals are asynchronous. However, the alignment by DTW is clearly not consistent to the assumption in Figure \ref{fig2-2}(b). This is a structural limitation of DTW because outputs from each signal are forced to be matched even if the signals are asynchronous. This is why the alignment errors of DTW alternate from $-0.2$ to $0.2$ in Figure \ref{fig2-2}(c). An analogous but more extreme case is shown in Figure S4 and S5 in the supplementary notes.

\subsubsection{Example 2: Query Signals Contaminated with a Small Number of Outliers}\label{sec2-3-2}

\begin{figure}
\centering
\includegraphics[width=1\textwidth]{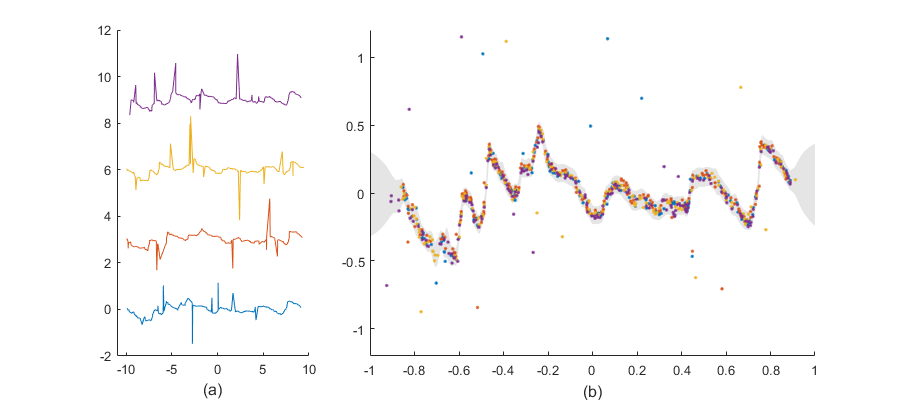}
\caption{(a) Four contaminated signals obtained by respectively perturbing inputs of the same sequence. (b) Medians of projected signals on the constructed profile. The shaded region depicts the 95\% confidence band of the SA-GPR profile.}
\label{fig2-5}
\end{figure}

It is natural to observe outliers in the real data. Although outliers can be discarded in the preprocessing step, results from the alignment algorithm should ideally be robust to them. A problem is that the criteria used to identify outliers are essentially subjective. One advantage of the profile-dependent algorithms is that outliers can be easily and rigorously defined as data that do not follow the profile. Figure \ref{fig2-5} represents the robustness of our SA-GPR on the outliers.

\subsubsection{Example 3: Heteroscedastic Noises}\label{sec2-3-3}

\begin{figure}
\centering
\includegraphics[width=1\textwidth]{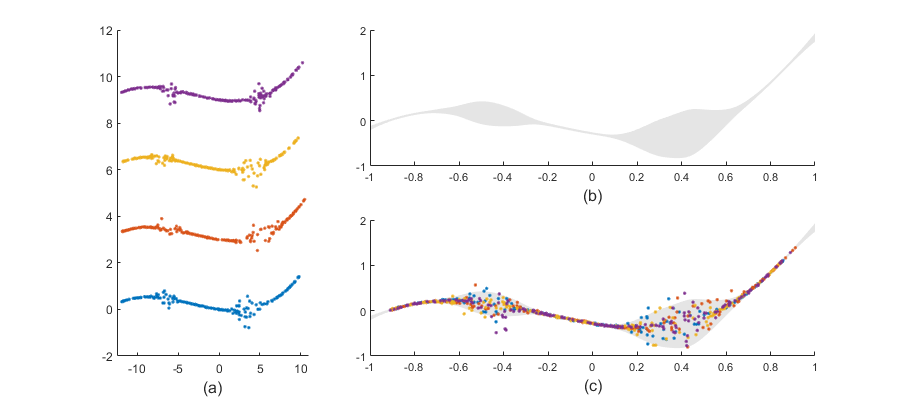}
\caption{(a) Four heteroscedastic signals obtained by respectively perturbing inputs of the same sequence. (b) SA-GPR profile constructed from the four signals in (a), where the shaded region depicts its 95\% confidence band. (c) Medians of projected signals on the constructed profile.}
\label{fig2-6}
\end{figure}

Real data may be noisy. If noise is homoscedastic, defining consistent penalties over inputs in DTW-based methods is straightforward. However, if noise is heteroscedastic, consistent penalty assignments would exaggerate mismatches at regions in which noises have high variances. Also, in some cases it is more important to capture the heteroscedasticity over inputs rather than to match signals exactly. With a heteroscedastic profile, mismatches to the profile are not penalized if the variances from the profile are high. Figure \ref{fig2-6} shows an example of such cases, with the heteroscedasticity inferred by the method in \cite{Lee2019}.

\subsubsection{Example 4: Signals Not Completely Overlapping One Another}\label{sec2-3-4}

\begin{figure}
\centering
\includegraphics[width=1\textwidth]{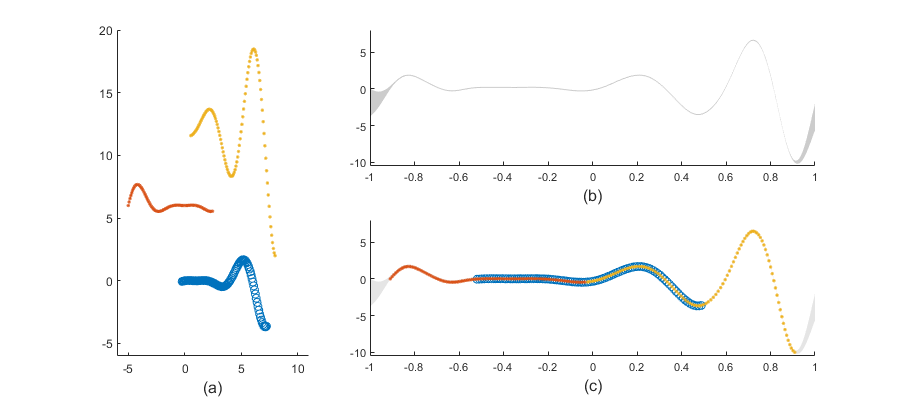}
\caption{(a) Three signals to be concatenated. (b) SA-GPR profile constructed from the three signals in (a), where the shaded region depicts its 95\% confidence band. (c) Medians of the projected signals on the constructed profile.}
\label{fig2-7}
\end{figure}

A profile is designed to cover all signals, and conversely, our profile construction algorithm is designed to make each signal contribute to the profile, whether or not it covers the entire profile. Figure \ref{fig2-7} shows an example of three signals aligned by SA-GPR where there is no overlap.

\section{BIGMACS: Bayesian Inference Gaussian Process Multiproxy Algorithm of Continuous Signals}\label{sec3}

The Bayesian Inference Gaussian Process Multiproxy Algorithm of Continuous Signals (BIGMACS) [\url{http://github.com/eilion/BIGMACS}] is a software to construct ocean sediment core age models from their radiocarbon and benthic ${\delta}^{18}{\rm O}$ proxies, where the former is an dating proxy and the latter is a synchronizing proxy. BIGMACS can also construct a profile, or stack in a paleoceanographic term, from multiple sediment cores. BIGMACS is an application of the SA-GPR to paleoceanography.

\subsection{The Problem and Terminology}\label{sec3-1}

In the field of paleoceanography, a well-established method for dating ocean sediment cores is based on the ratio of oxygen isotopes in foraminiferal shells. Specifically, ${\delta}^{18}{\rm O}$ is the ratio of stable isotopes ${}^{18}{\rm O}$ and ${}^{16}{\rm O}$ relative to a laboratory standard. The ${\delta}^{18}{\rm O}$ of foraminiferal shells is a common proxy for water temperature, salinity, and global ice volume. The ${\delta}^{18}{\rm O}$ of benthic foraminifera, which live on the seafloor, is often used as a global climate parameter because global ice volume accounts for approximately half the variance through time \cite{Spratt2016} while deep-water temperature and salinity have relatively little spatial variability.

An age model is a function mapping depth in a sediment core to age. The age model for an input core which lacks an dating proxy can be estimated by aligning its benthic ${\delta}^{18}{\rm O}$ records to a target record that has an age model. In this process, the input core can adopt the age model of the target indirectly. The target record is often an average of multiple cores, such as the LR04 stack \cite{Lisiecki2005}. The target can also be a probabilistic model developed from multiple records (Prob-Stack \cite{Ahn2016}). Here, the term “stack” replaces the “profile” in statistics.

Although benthic ${\delta}^{18}{\rm O}$ is a global parameter, its local variability between core locations can cause significant uncertainty in the age inference when studying millennial scale events. Benthic ${\delta}^{18}{\rm O}$ age models can result in age model errors up to 4 kiloyears \cite{Skinner2005,Stern2014} due to local effects that might corrupt the inference. For example, after the Last Glacial Maximum, 19-23 kiloyears ago, melting ice sheets changed temperature and salinity gradients in the deep ocean and altered ocean circulation \cite{Adkins2002,McManus2004}. Asynchronous temperature changes and ocean circulation rates caused by ice melt are thought to influence benthic ${\delta}^{18}{\rm O}$ records from 18 to 11 kiloyears ago and could cause bias in ${\delta}^{18}{\rm O}$-aligned age models [e.g., \cite{Waelbroeck2011,Gebbie2012}].

Layers of the sediment cores can be directly dated with radiocarbon ages. However, this dating method is limited to only the most recent 50 kiloyears due to radioactive decay, whereas benthic ${\delta}^{18}{\rm O}$ records from ocean sediment cores extend to more than 60 million years ago. Furthermore, disturbances to sediments deposited on the seafloor can sometimes result in age reversals within the sequence of sediment recovered by coring. Additionally, the resolution of radiocarbon measurements is frequently lower than that of ${\delta}^{18}{\rm O}$. As a result, age inferences of cores with only radiocarbon data of low resolution are strongly dependent on the assumptions regarding variability in the rate of sediment accumulation in the intervals between data points.

Therefore, utilizing both age and synchronizing proxies is desirable to improve age inferences for ocean sediment cores. Dating proxies from a set of cores are integrated according to the alignments based on the synchronizing proxies, and local variability of synchronizing proxies between core locations is accessed by the dating proxies. BIGMACS is designed to exploit both types of proxies simultaneously. Moreover, it continuously samples ages for layer depths (positions) and their proxy observations in the core. One more advantage is that BIGMACS can construct a continuous stack, regardless of how many cores are available, how high the resolution of each core is, and how irregular the proxy observations are.

\subsection{Previous Approaches}\label{sec3-2}

Radiocarbon measurements from specific depths within a core are converted into calendar (real) ages using calibration curves \cite{reimer2020,hogg2020,Heaton2020} that account for the rate of radioactive decay and past changes of atmospheric radiocarbon production rates. Thus, radiocarbon proxies allow us to access calendar ages corresponding to sediment layer depths in a core. \cite{Christen2009} developed an algorithm to construct sediment core age models from radiocarbon data based on a generalized Student’s t-distribution that is robust to outliers. This algorithm includes the uncertainties from radiocarbon measurements and tuning hyperparameters that reflect the ${}^{14}{\rm C}$ composition of seawater relative to the atmosphere, known as the reservoir age.

Multiple studies have constructed dynamic models that simulate changes in sediment accumulation rates, defined as the ratio of depth to age increments). \cite{Blaauw2005} adopt a piecewise linear approach with automatic section selection and impose constraints on accumulation rates whereas \cite{Haslett2008} use a bivariate monotone Markov process with gamma increments. \cite{Blaauw2011} developed an algorithm, called Bacon (Bayesian Accumulation), to construct an age-depth model by adopting an autoregressive gamma process for accumulation rates and a Student’s t-distribution for radiocarbon proxy. An adaptive MCMC algorithm is implemented to sample ages.

Synchronizing age assignments depend on record alignments and allow a core to utilize ages from a different core or stack that has dating proxies. A deterministic alignment algorithm, Match, was developed using dynamic programming \cite{Lisiecki2002}. \cite{Lisiecki2005} used Match to align benthic ${\delta}^{18}{\rm O}$ data from 57 globally distributed deep-sea sediment cores. These data were averaged to calculate a stack, called LR04, which is commonly used as a standard reference for benthic ${\delta}^{18}{\rm O}$ change over the past 5.3 million years.

Ages can be inferred by alignment to the LR04 stack, similar to the way profile hidden Markov models (HMMs) can be employed in biological sequence alignments (details can be found in \cite{Durbin1998}). \cite{Lin2014} tackled the age assignments problem using a probabilistic alignment model called HMM-Match. The HMM-Match emission model for ${\delta}^{18}{\rm O}$ data is based on Gaussian distributions with time varying mean ${\delta}^{18}{\rm O}$ values from LR04 and a constant core-dependent standard deviation learned by the Baum-Welch EM algorithm. The transition model accounts for the probability distribution of accumulation rates using a log-normal mixture based on radiocarbon observations from 37 cores. \cite{Ahn2016} constructed a stack (named Prob-stack) from 180 globally distributed benthic ${\delta}^{18}{\rm O}$ cores with an algorithm based on HMM-Match, called HMM-Stack. Each point in Prob-stack is described by a Gaussian distribution of ${\delta}^{18}{\rm O}$ that varies along the core.

HMMs define inferred ages on discrete spaces. This is problematic when the input core has a higher resolution than the target stack or when a proportional accumulation model is employed like the one used in Match, in HMM-Match, and BIGMACS. Increasing the resolution of the target stack cannot be an ultimate solution because the time complexity of an HMM is quadratic to the size of its hidden space, so it soon becomes infeasible as the resolution increases. Because BIGMACS samples ages continuously by a hybrid of the particle smoothing and Metropolis-Hastings, it does not suffer from the same drawback as models based on the HMM.

\subsection{Models}\label{sec3-3}

BIGMACS is an application of SA-GPR, thus it is of the same structure but defines specific emission and transitoin models for the application to ocean sediment cores. Here, the space $\mathcal{X}$ on which the depths (positions) are defined is $\mathbb{R}$ and $\mathcal{Y}$ for the proxy observations is $\mathbb{R}^2$. Alignments are now ages, thus defined on $\mathcal{S}=\left[0,{\rm{S}}_{max}\right]$, where ${\rm{S}}_{max}$ is the maximum age of the cores.

The transition model (prior of the alignments) of BIGMACS is more complicated than SA-GPR and defined backwardly, to reflect the sediment accumulation model more rigorously. For each depth $n\in\left\{1,2,\cdots,{\rm{N}}_m\right\}$ in core $m$, we define a medium latent variable ${\rm{W}}^{\left(m\right)}=\left\{{\rm{W}}_n^{\left(m\right)}\right\}_{n=1}^{{\rm{N}}_m}$ where each ${\rm{W}}_n^{\left(m\right)}\in\left\{\mathbb{C},\mathbb{A},\mathbb{E}\right\}$ that stands for contraction, average, and expansion, respectively. Then, the transition model is given as follows:
\begin{equation}
\pi_1\left({\rm{W}}_n^{\left(m\right)}\middle|{\rm{W}}_{n+1}^{\left(m\right)},\phi^{\left(m\right)}\right)=\phi_{{\rm{W}}_{n+1}^{\left(m\right)},{\rm{W}}_n^{\left(m\right)}}^{\left(m\right)},
\label{equ3-3-1}
\end{equation}
\begin{equation}
\begin{aligned}
& \pi_2\left({\rm{Z}}_n^{\left(m\right)}\middle|{\rm{Z}}_{n+1}^{\left(m\right)},{\rm{W}}_n^{\left(m\right)},x_n^{\left(m\right)},x_{n+1}^{\left(m\right)},r^{\left(m\right)}\right) \\ & \propto {\rm{Gamma}}\left(\frac{{\rm{Z}}_{n+1}^{\left(m\right)}-{\rm{Z}}_n^{\left(m\right)}}{r^{\left(m\right)}\left(x_{n+1}^{\left(m\right)}-x_n^{\left(m\right)}\right)}\middle|\alpha,\beta\right) \cdot 1_{\left\{\frac{{\rm{Z}}_{n+1}^{\left(m\right)}-{\rm{Z}}_n^{\left(m\right)}}{r^{\left(m\right)}\left(x_{n+1}^{\left(m\right)}-x_n^{\left(m\right)}\right)} \in {\rm{I}}_{{\rm{W}}_n^{\left(m\right)}} \right\}} \left( {\rm{Z}}_n^{\left(m\right)} \right),
\end{aligned}
\label{equ3-3-2}
\end{equation}
where the last term in (\ref{equ3-3-2}) assumes that ${\rm{I}}_\mathbb{C}=\left(0,0.9220\right)$, ${\rm{I}}_\mathbb{A}=\left[\left.0.9220,1.0850\right)\right.$, ${\rm{I}}_\mathbb{E}=\left[\left.1.0850,\infty\right)\right.$. In words, each latent variable ${\rm{W}}_n^{\left(m\right)}$ confines the transition from an age ${\rm{Z}}_{n+1}^{\left(m\right)}$ to another ${\rm{Z}}_n^{\left(m\right)}$ in one of the three regions $\left\{\mathbb{C},\mathbb{A},\mathbb{E}\right\}$ that correspond to ${\rm{I}}_\mathbb{C}$, ${\rm{I}}_\mathbb{A}$ and ${\rm{I}}_\mathbb{E}$. $r^{\left(m\right)}$ is a depth-scale parameter that rescales ${\rm{X}}^{\left(m\right)}$ to adjust for differences in average accumulation rates. A transition matrix, $\phi^{\left(m\right)}$, maps $\left\{\mathbb{C},\mathbb{A},\mathbb{E}\right\}$ to itself.

The transition model is an ${\rm{AR}}\left(2\right)$ because the previous accumulation rate from ${\rm{Z}}_{n+2}^{\left(m\right)}$ to ${\rm{Z}}_{n+1}^{\left(m\right)}$ that is stored at ${\rm{W}}_{n+1}^{\left(m\right)}$ affects the choice of the current ${\rm{W}}_n^{\left(m\right)}$. This in turn influences the current accumulation rate from ${\rm{Z}}_{n+1}^{\left(m\right)}$ to ${\rm{Z}}_n^{\left(m\right)}$. The hyperparameters $\alpha$ and $\beta$ are fixed in the procedure to avoid overfitting; their values are pre-learned from the same radiocarbon dataset for learning the log-normal mixture model in \cite{Lin2014}.

From (\ref{equ3-3-1}) and (\ref{equ3-3-2}), the full transition model is given as follows:
\begin{equation}
\begin{aligned}
& \pi\left({\rm{Z}}_n^{\left(m\right)},{\rm{W}}_n^{\left(m\right)}\middle|{\rm{Z}}_{n+1}^{\left(m\right)},{\rm{W}}_{n+1}^{\left(m\right)},x_n^{\left(m\right)},x_{n+1}^{\left(m\right)},\phi^{\left(m\right)},r^{\left(m\right)}\right) \\ & \propto \phi_{{\rm{W}}_{n+1}^{\left(m\right)},{\rm{W}}_n^{\left(m\right)}}^{\left(m\right)} \cdot  {\rm{Gamma}}\left(\frac{{\rm{Z}}_{n+1}^{\left(m\right)}-{\rm{Z}}_n^{\left(m\right)}}{r^{\left(m\right)}\left(x_{n+1}^{\left(m\right)}-x_n^{\left(m\right)}\right)}\middle|\alpha,\beta\right)  \cdot 1_{\left\{\frac{{\rm{Z}}_{n+1}^{\left(m\right)}-{\rm{Z}}_n^{\left(m\right)}}{r^{\left(m\right)}\left(x_{n+1}^{\left(m\right)}-x_n^{\left(m\right)}\right)}\in {\rm{I}}_{{\rm{W}}_n^{\left(m\right)}}  \right\}} \left( {\rm{Z}}_n^{\left(m\right)} \right).
\end{aligned}
\label{equ3-3-3}
\end{equation}

At each position (depth), we have two types of observations: radiocarbon and benthic ${\delta}^{18}{\rm O}$. For radiocarbon proxy, we adopt a model of \cite{Christen2009}. For each depth, the emission model for radiocarbon is given as follows:
\begin{equation}
g_1\left(y_{n,1}^{\left(m\right)}\middle|\varphi_1\left({\rm{Z}}_n^{\left(m\right)}\right)\right)=\mathcal{T}_{2a_1}\left(y_{n,1}^{\left(m\right)}\middle|\mu_{\rm{C}}\left({\rm{Z}}_n^{\left(m\right)}\right)+\varrho_n^{\left(m\right)},\sqrt{\frac{b_1}{a_1}\left(\sigma_{\rm{C}}^2\left({\rm{Z}}_n^{\left(m\right)}\right)+\varsigma_n^{\left(m\right)}\right)}\right),
\label{equ3-3-4}
\end{equation}
where $\varrho_n^{\left(m\right)}$ and $\varsigma_n^{\left(m\right)}$ are given together with the radiocarbon determination (a measurement of the amount of radiocarbon in a sample) $y_{n,1}^{\left(m\right)}$ a priori, and $\mu_C$ and $\sigma_C$ are the mean and standard deviation functions of the radiocarbon calibration curve \cite{reimer2020,hogg2020,Heaton2020}. In words, (\ref{equ3-3-4}) translates the calendar age ${\rm{Z}}_n^{\left(m\right)}$ to the radiocarbon determination $y_{n,1}^{\left(m\right)}$. $a_{1}=3$ and $b_{1}=4$ are the fixed hyperparameters, just as \cite{Christen2009} adopt. Though SA-GPR also allows to construct the profile for radiocarbon calibration curve, here we assume that it is given a priori and not to be updated.

For benthic ${\delta}^{18}{\rm O}$, the emission model basically follows the stack, but in a different form to deal with potential outliers, as follows:
\begin{equation}
\begin{aligned}
& g_2\left(y_{n,2}^{\left(m\right)}\middle|\varphi_2\left({\rm{Z}}_n^{\left(m\right)},\sigma^{\left(m\right)},h^{\left(m\right)}\right)\right) \\ & =\mathcal{T}_{2{a_{2}}}\left(y_{n,2}^{\left(m\right)}\middle|\sigma^{\left(m\right)}\overline{\mu}\left({\rm{Z}}_n^{\left(m\right)}\right)+h^{\left(m\right)},\sqrt{\frac{b_2}{a_2}\left(\sigma^{\left(m\right)}\right)^2\overline{\nu}\left({\rm{Z}}_n^{\left(m\right)}\right)}\right),
\end{aligned}
\label{equ3-3-5}
\end{equation}
\begin{equation}
\log{\pi_{em}\left(\sigma^{\left(m\right)},h^{\left(m\right)}\right)}\propto-\frac{1}{2}\left(\frac{h^{\left(m\right)}-\underline{h}}{\underline{\sigma}}\right)^2-2\left(\underline{\alpha}+1\right)\log{\sigma^{\left(m\right)}}-\underline{\beta}\left(\sigma^{\left(m\right)}\right)^{-2},
\label{equ3-3-6}
\end{equation}
where $\overline{\mu}$ and $\overline{\nu}$ are given by the ${\delta}^{18}{\rm O}$ stack and $\sigma^{\left(m\right)}$ and $h^{\left(m\right)}$ are the core-specific scale and shift parameters, respectively. $a_{2}=3$, $b_{2}=4$, $\underline{h}=0$, $\underline{\sigma}=1$ and $\underline{\alpha}=\underline{\beta}=1$ are the fixed hyperparameters.

We assume that radiocarbon and benthic ${\delta}^{18}{\rm O}$ proxies are conditionally independent given the alignment (age). This assumption is reasonable because they are measured from the fossils of different species (which lived at different depths in the ocean). From (\ref{equ3-3-4}) and (\ref{equ3-3-5}), the full emission model is given as follows:
\begin{equation}
\begin{aligned}
& g\left(y_{n,1}^{\left(m\right)},y_{n,2}^{\left(m\right)}\middle|\varphi\left({\rm{Z}}_n^{\left(m\right)},\sigma^{\left(m\right)},h^{\left(m\right)}\right)\right) \\ & =\mathcal{T}_{2a_1}\left(y_{n,1}^{\left(m\right)}\middle|\mu_C\left({\rm{Z}}_n^{\left(m\right)}\right)+\varrho_n^{\left(m\right)},\sqrt{\frac{b_1}{a_1}\left(\sigma_C^2\left({\rm{Z}}_n^{\left(m\right)}\right)+\varsigma_n^{\left(m\right)}\right)}\right) \\ & \times \mathcal{T}_{2a_2}\left(y_{n,2}^{\left(m\right)}\middle|\sigma^{\left(m\right)}\overline{\mu}\left({\rm{Z}}_n^{\left(m\right)}\right)+h^{\left(m\right)},\sqrt{\frac{b_2}{a_2}\left(\sigma^{\left(m\right)}\right)^2\overline{\nu}\left({\rm{Z}}_n^{\left(m\right)}\right)}\right).
\end{aligned}
\label{equ3-3-7}
\end{equation}

In practice, we have \emph{at most} two types of observations at each depth, although it is quite rare to observe both proxies at one depth. Additionally, multiple observations of one type of proxy are sometimes made at a given depth. Therefore, the emission model (\ref{equ3-3-7}) depends on the number and types of observations at each depth. Also notice that this formulation can be extended to any tuples of different proxies as long as they are believed to be conditionally independent given ages, for example, tephra layers.

For profile (stack) construction, we apply the GPR with OU kernel \cite{Rasmussen2005}. It is a special case of the Matérn class of covariance functions when the degree of differentiability is 0.5 and gives rise to a particular form of an ${\rm{AR}}\left(1\right)$ model \cite{Roberts2013}. We applied the heteroscedastic GPR to the stack construction to treat the heteroscedastic variance of benthic ${\delta}^{18}{\rm O}$ over ages.

\section{Application: Dual Proxy Stack Construction for Paleoceanography}\label{sec4}

Although benthic ${\delta}^{18}{\rm O}$ is usually interpreted as a global parameter, local-to-regional differences can occur when climate change occurs on the same timescale as the 1-2 kiloyear mixing time of the deep ocean. For example, rapid melting of ice sheets following the Last Glacial Maximum (19-23 kiloyears ago) produced regional variations in ${\delta}^{18}{\rm O}$ as the melt water and temperature change signals propagated initially from deep water formation regions to the rest of the ocean’s interior \cite{Waelbroeck2011}. Because this process can cause differences of up to 4 kiloyears in the age of ${\delta}^{18}{\rm O}$ changes between some regions \cite{Skinner2005,Stern2014}, it is important to construct local stacks from (small) sets of cores that are believed to share the local effects; recall that a stack summarizes the ${\delta}^{18}{\rm O}$ pattern of cores. Here, we define a set of cores as homogeneous if they share the same minor effects (e.g., sampling well-mixed seawater originated from same source locations at similar times); mathematically, no structurally correlated variation in means and variances is expected across the cores. We assume that homogeneous cores can be simultaneously explained by the same parameters, so it makes sense to construct a stack for them. A local stack of a set of homogeneous cores is provided by the Gaussian model of ${\delta}^{18}{\rm O}$ values in those cores.

BIGMACS can align the cores and construct their stack, regardless of how many cores are given. Here we construct a local stack, which we call the deep northeastern Atlantic (DNEA) stack, using five cores containing both radiocarbon and benthic ${\delta}^{18}{\rm O}$. The availability of radiocarbon allows direct access to the calendar ages to enhance the age inferences; this is essentially useful in the Holocene (0-11.7 kiloyears ago) where the ${\delta}^{18}{\rm O}$ signal-to-noise ratio is low. We assume seawater properties at these sites are sufficiently similar to be considered homogeneous. Two of these cores are from the Iberian margin (MD95-2042 and MD99-2334), and three are from the northwest African continental slope (GeoB7920-5, GeoB9508-5 and GeoB9526-5), as shown in figure S6 in the supplementary notes. These cores come from sites bathed today by approximately 75\% Northern Sourced Water (NSW) and 25\% Southern Sourced Water (SSW); details can be found in Section S1 of the supplementary notes. A regional Deep North Atlantic (DNA) stack from \cite{Lisiecki2016} was used to initialize the iterative algorithm. After presenting the DNEA local stack, we provide two examples of using BIGMACS to generate age estimates for other nearby cores by aligning them to the stack.

\begin{figure}
\centering
\includegraphics[width=0.85\textwidth]{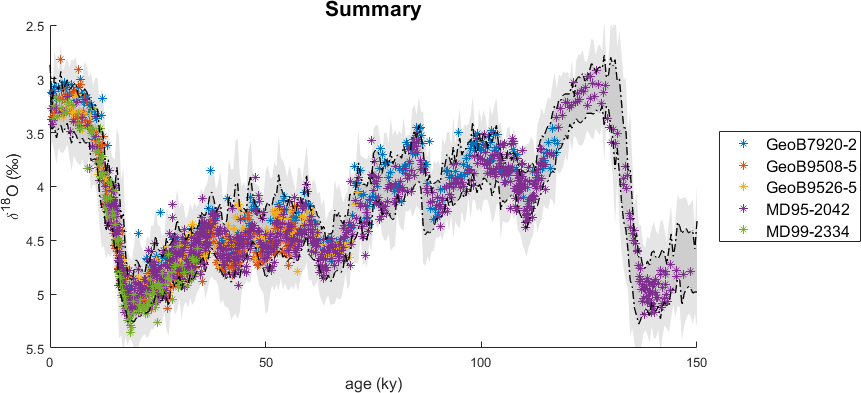}
\includegraphics[width=0.85\textwidth]{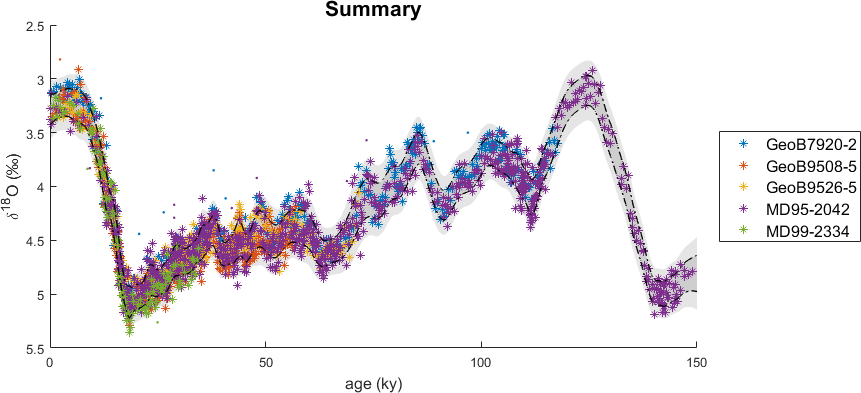}
\includegraphics[width=0.85\textwidth]{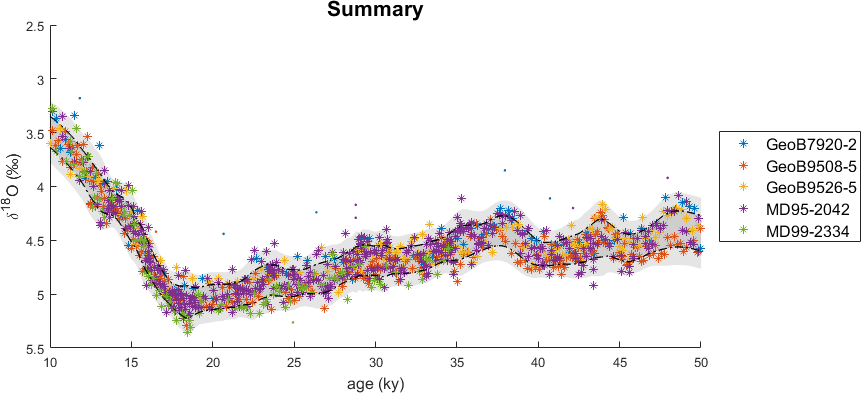}
\caption{Core alignments for stacking. The upper panel shows those to the DNA stack, while the middle panel is our dual-proxy DNEA stack. Stars indicate medians of data classified as inliers and dots represent outliers, after translations. The darker and brighter gray regions are the 1-sigma and 2-sigma of the stacks, respectively. The dot-dash lines indicate their boundary. The bottom panel is a portion of the DNEA stack at 10-50 kiloyears.}
\label{fig4-2}
\end{figure}

Figure \ref{fig4-2} compares the DNEA stack we constructed here to the regional DNA stack from \cite{Lisiecki2016}. Variability in benthic ${\delta}^{18}{\rm O}$ in the regional DNA stack is considerably larger than in the local DNEA stack. Note that most of the DNEA medians are inside the 1-sigma of the DNA stack, which is expected to contain 68\% of them. Higher variances in the DNA stack compared to the DNEA stack may stem from benthic ${\delta}^{18}{\rm O}$ differences within the broader North Atlantic region, record-specific mean shifts applied to the DNEA stack but not the construction of the DNA stack, and/or the discrete nature of the algorithm used to construct the DNA stack. Another contributing factor to the tighter DNEA stack variance is the automatic detection and removal of the outlying observations. The tighter variance contributes to less uncertainty in sediment core age estimates inferred from this stack.

The smoothness of the local stack stems from the fact that the GP model captures correlations between all the data points with a heavier weight placed on near neighbors, thus limiting sudden large changes. In this analysis we employed the OU kernel, whose correlation function decreases more quickly with distance than most other kernels. Nevertheless, the resulting dual proxy stack is smoother than the DNA stack and still captures well-known millennial-scale climatic events. For example, figure \ref{fig4-2}(c) shows four peaks at 24, 29, 38 and 46 kiloyears ago, which correspond to the Heinrich events H2 to H5 \cite{Skinner2013,Lisiecki2016}. The ability to resolve such short-lived features will improve the accuracy of age estimations for cores with high-resolution ${\delta}^{18}{\rm O}$ records.

\begin{figure}
\centering
\includegraphics[width=0.85\textwidth]{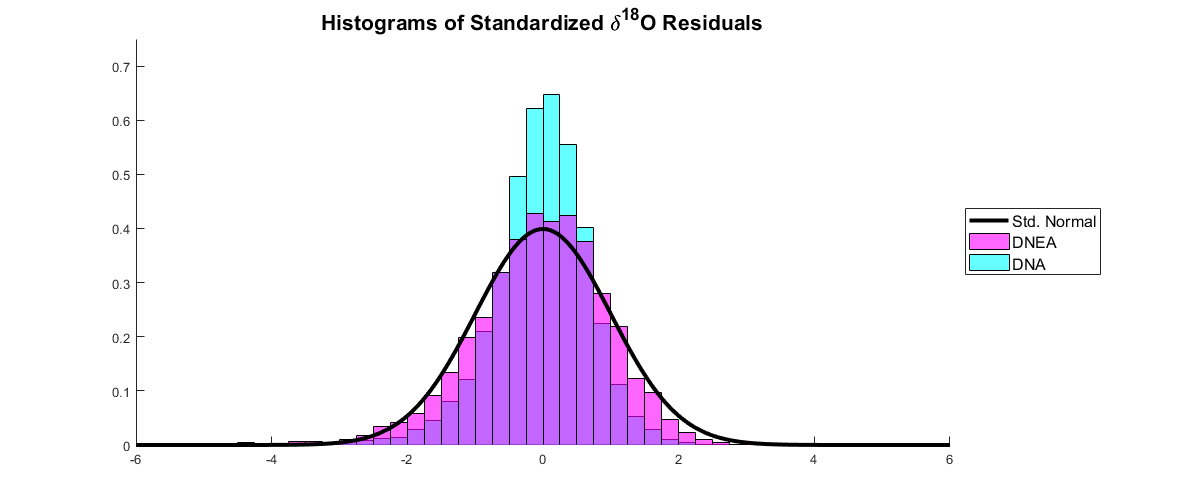}
\caption{Histograms of ${\delta}^{18}{\rm O}$ to our DNEA stack and the initial DNA stack. Each ${\delta}^{18}{\rm O}$ is standardized by the mean and standard deviation of the stack at the median of its sampled ages.}
\label{fig4-8}
\end{figure}

\begin{figure}
\centering
\includegraphics[width=1\textwidth]{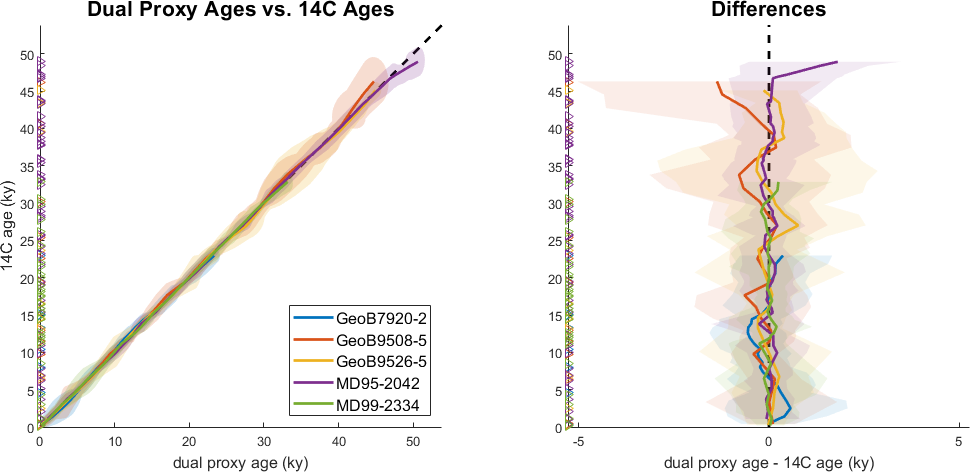}
\caption{The comparisons between inferred dual proxy ages and radiocarbon ages of records. The left and right panels show the comparison and differences, respectively. Shaded regions indicate 95\% confidence regions and the black dashed line is just a diagonal.}
\label{fig4-3}
\end{figure}

Figure \ref{fig4-8} shows the histograms of standardized ${\delta}^{18}{\rm O}$ from the six records in the dual-proxy stack with respect to the DNA stack and the new DNEA stack. The existence of only a small departure from the standard normal distribution to the DNEA stack supports the validity of a Gaussian model and our choice of homoscedastic GP regression in the stack construction. Figure \ref{fig4-3} compares ages inferred using both ${\delta}^{18}{\rm O}$ and radiocarbon (dual proxy) to those using radiocarbon only (analogous to Bacon \cite{Blaauw2011}). Overall agreement between cores (to within uncertainty) supports our assumption that benthic ${\delta}^{18}{\rm O}$ is synchronous and homogeneous among sites included in the local stack. Some departures from the diagonal are expected, considering influences from their ${\delta}^{18}{\rm O}$ data.

If the ${\delta}^{18}{\rm O}$ record for a particular core site is believed to be homogeneous to the ${\delta}^{18}{\rm O}$ in the local stack, its ages can be inferred indirectly by dual proxy alignment to the local stack or ${\delta}^{18}{\rm O}$-only alignment if radiocarbon data are unavailable. Here we present example alignments to the DNEA stack for two sediment cores from the same region (GIK13289-2 and SU90-08) which are assumed to share the same local ${\delta}^{18}{\rm O}$ signal as the stack. However, sometimes it may be difficult to ascertain whether proxy signals are homogeneous across two or more sites through time. When assessing whether core sites share the same local ${\delta}^{18}{\rm O}$ signal, one should consider not only whether the cores share the same water composition today but also how ocean circulation may have changed through time. We evaluate whether the cores used here are homogeneous in Section S1 of the supplementary notes.

\begin{figure}
\centering
\includegraphics[width=0.80\textwidth]{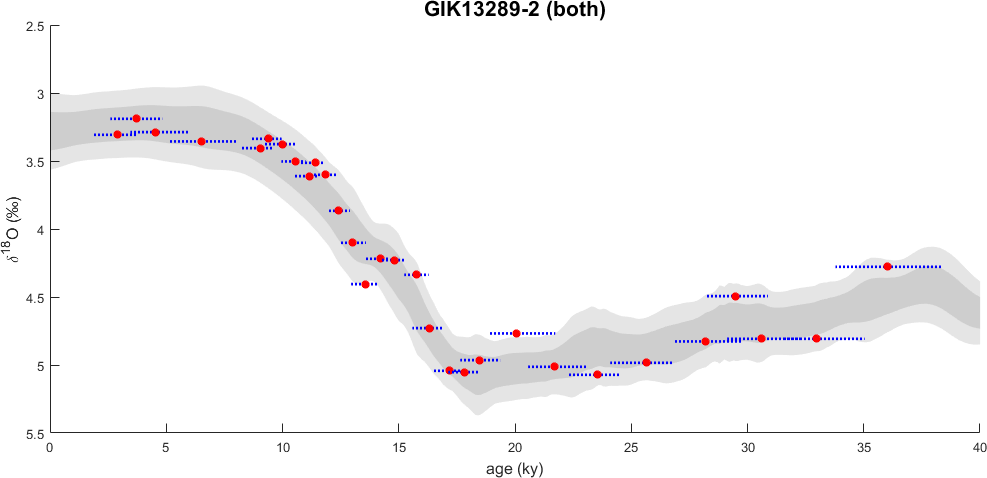}
\includegraphics[width=1.00\textwidth]{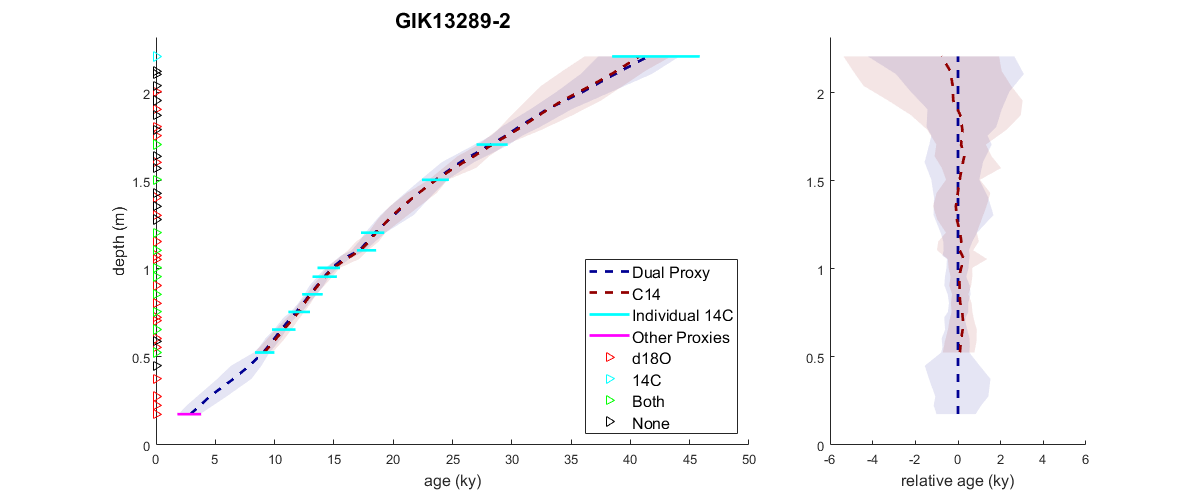}
\caption{Dual proxy age inferences and alignments of GIK13289-2 to DNEA stack. In the upper panel, the darker and brighter areas show the 1-sigma and 2-sigma of the stacks, red points and blue dotted bars indicate medians and 95\% confidence intervals of inlier age samples for ${\delta}^{18}{\rm O}$ data, blue points and red dotted bars are the outliers. In the left below panel, cyan and magenta bars indicate 95\% confidence intervals obtained independently from radiocarbon data and other proxies used in \cite{Sarnthein1994}, respectively, and blue and red areas show the 95\% confidence bands of age inferences from radiocarbon only and both proxies, respectively. In the right below panel, medians (dashed lines) and 95\% confidence intervals of relative ages to the medians of dual proxy ages are shown.}
\label{fig4-4}
\end{figure}

\begin{figure}
\centering
\includegraphics[width=0.80\textwidth]{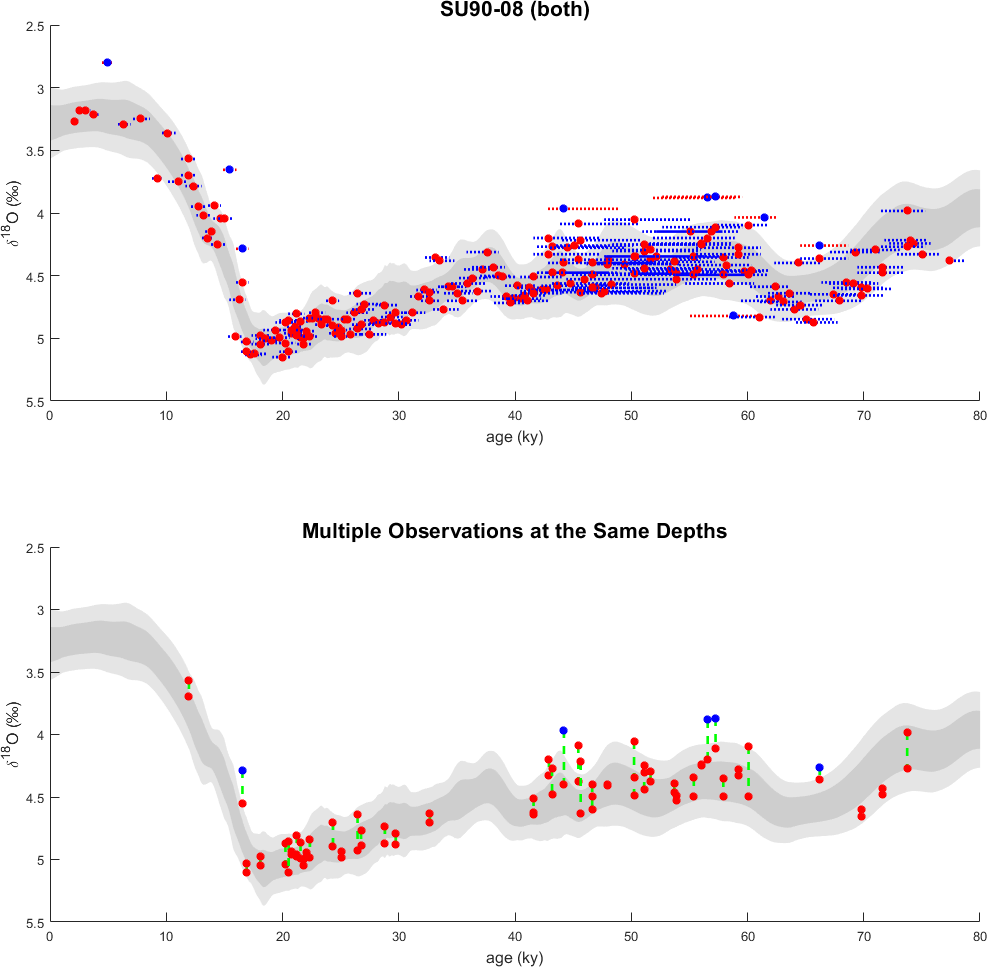}
\includegraphics[width=1.00\textwidth]{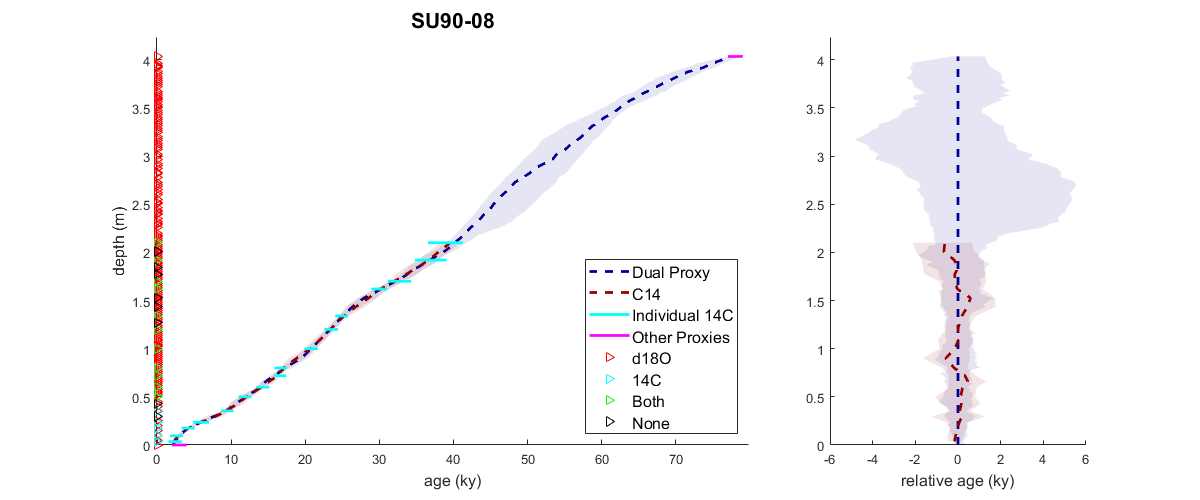}
\caption{Dual proxy age inferences and alignments of SU90-08 to DNEA stack. Similar format to Figure 8, but this figure also shows how multiple observations at the same depths are dealt with in the alignment algorithm. The magenta bars indicate 95\% confidence intervals obtained independently from other proxies used in \cite{Grousset1993}.}
\label{fig4-5}
\end{figure}

Figure \ref{fig4-4} and \ref{fig4-5} show the results of GIK13289-2 and SU90-08, respectively. The upper panels show that the translated and aligned ${\delta}^{18}{\rm O}$ data mainly fall within the stack’s confidence intervals. The lower left panels show that the inferred ages mainly pass through the confidence intervals from individual radiocarbon proxies (SU90-08 has radiocarbon data between 10-40 kiloyears only and beyond that ages are inferred only based on ${\delta}^{18}{\rm O}$ alignment to the stack). Because stacks with dual proxy ages have narrower confidence intervals than single proxy stacks, they are more informative for stack alignment and produce smaller age uncertainties. For example, compare figure \ref{fig4-4} and \ref{fig4-5} with S7 and S8 (in the supplementary notes) that show alignments to the DNA ${\delta}^{18}{\rm O}$ stack.

\section{Discussion}\label{sec5}

SA-GPR shares the advantages of the other profile-dependent multiple signal alignment algorithms. Aligning signals not only finds the best matches between observations in signals but also integrates the fragmented information stored in each signal into the profile. It is flexible in the choice of transition and emission models, so one can prevent the reversal problem; the order of original inputs should be preserved in the alignments. Once the profile is constructed from a set of signals, it can be used as an alignment target for new signals. These advantages allow age and synchronizing proxies to be exploited more efficiently to infer the ages of ocean sediment cores continuously, as seen in the BIGMACS results of section \ref{sec4}.

Our profile construction method is constructive rather than implicit, so the obtained profiles are intuitive and easy to interpret. Once SA-GPR constructs a profile, it can be understood as the shared pattern within uncertainty of the query signals. This means that the profile can be utilized as a representation of the query signals for other meaningful purposes. For example, the mean function of the DNEA stack in section \ref{sec4} forms a global parameter directly used for inferring past climate changes.

The combination of the particle smoothing algorithm and Metropolis-Hastings algorithm is guaranteed to sample alignments given positions and observations continuously. Note that the particle smoothing is applied only for the initialization. In SA-GPR, there are several reasons why HMM is not considered in the initialization. First, if the dynamics of a latent series do not allow “stays” (i.e., alignments must be strictly increasing), then the downgraded transition model in the HMM cannot perfectly reflect it whereas the particle smoothing is well suited to this task. Second, the particle smoothing can focus only on each latent variable in possible ranges of the profile while HMM considers all candidates for assigning it, which make the algorithm more efficient.

Even if continuous sampling is available, it will soon become insignificant without a continuously defined profile. SA-GPR, which depends on the Gaussian process regression (GPR) is superior to the algorithms based on simple interpolations because of automatic accounting for greater uncertainty in interpolation when neighboring points are more widely spread, which is clearly desirable when unobserved outputs are filled in; see the connection between Matérn kernels and continuous AR models in \cite{Rasmussen2005}. The desirable feature assumes that the closer the query position is to the data, the higher the associated observation is correlated with them. The strength of this effect varies with the kernel covariance functions selected. Moreover, GPR is more robust to unexpected changes in the pattern of the data because it is nonparametric.

The time complexity of the profile construction and signal alignment algorithms of SA-GPR are $\mathcal{O}\left(\left|{\rm{Y}}\right|\left|\underline{\rm{Z}}\right|^2{\rm{L}}\right)$ and $\mathcal{O}\left(\left|{\rm{Y}}\right|{\rm{P}}^2{\rm{L}}\right)$, respectively, where $\underline{\rm{Z}}$ is the pseudo-input in (\ref{equ2-1-8}), $\rm{P}$ is the number of particles in the particle smoothing algorithm, and $\rm{L}$ is the number of sampled alignments. That both time complexities are linear to the number of observations $\left|{\rm{Y}}\right|$ is a clear advantage over the other profile-free alignment methods that are based on the pairwise alignments, whose time complexities are structurally quadratic to $\left|{\rm{Y}}\right|$. Also, both aligning and profile construction steps can be parallelized over signals and sampled alignments, respectively.

Both the particle smoothing and Gaussian process regressions are sensitive to an issue that is relevant to the dimension of the alignments, “curse of dimensionality". However, SA-GPR is free from that tricky issue because here we consider the methods to be defined over a single ordered field $\mathcal{S}$, which is often a bounded interval in $\mathbb{R}$. Finding “best” kernel covariance function is still an issue to address, which greatly affects the performance of the profile construction. Domain-specific knowledge, such as that supplied by geoscientists, is highly valuable in making this choice.

\section{Conclusion}\label{sec6}

Here we present a novel multiple signal alignment framework, SA-GPR. Continuous alignments are performed on the given profile and continuous profiles are constructed from a limited set of signals. The combination of the particle smoothing and Metropolis-Hastings algorithms returns a set of continuously sampled alignments given the data and profile, as opposed to only one optimal alignment for each signal. The profile construction algorithm based on the Gaussian process regression returns a continuous profile regardless of the number of query signals. Our SA-GPR shares all the strengths of the existing alignment algorithms that depend on profiles but is more exact in the sense that profiles do not need to be discretized as sequential bins. The uncertainty of performance over the resolution of such bins is thereby eliminated. Future studies should attempt to reduce GP-specific limitations, such as kernel selection problems.

We then apply the SA-GPR algorithm to paleoceanography by developing the software, BIGMACS, which we use to construct a stack from five cores in the Deep Northeast Atlantic Ocean. To the best of our knowledge, this is the first probabilistic algorithm to align ocean sediment cores in continuous time and to include both age and synchronizing proxies in the alignment process. In addition, the algorithm used to construct this stack removes outliers in benthic ${\delta}^{18}{\rm O}$ data based on standardized objective criteria. The stack itself is a Gaussian process regression model, which we developed to address limitations in previous alignments of benthic ${\delta}^{18}{\rm O}$ observations to discrete stacks. Also, to capitalize on the continuous stack, we supplemented the particle smoothing used in the alignment step with a continuous time MCMC alignment procedure, which uses a continuous time sediment accumulation probability distribution. The stack, which includes radiocarbon data from every input record, is a better alignment target than a single record. Additionally, benthic ${\delta}^{18}{\rm O}$ data, which are often higher resolution then radiocarbon data, can improve calendar age inferences of the stack between radiocarbon ages, by reducing dependence on an assumed accumulation model. Lastly, other sediment cores can use the multi-proxy stack as an alignment target from which ages can be indirectly inferred.

Although here we present an application of SA-GPR to paleoceanography only, this framework can be applied to any field that requires the alignment of multiple signals, sequences, time series, or other equivalents. Their shared pattern can be integrated in the form of a profile which characterizes differences from the normal. Regression and prediction can be done for each signal based on its alignment to the others. Potential applications in other fields of study awaits further work in the future.

\section{Supplementary Notes: Assessing Homogeneity}\label{S1}

Sites may share a homogeneous signal if they are bathed by the same water mass. All of the cores we analyzed come from sites bathed today by approximately 75\% Northern Sourced Water (NSW, specifically North Atlantic Deep Water) and 25\% Southern Sourced Water (SSW, specifically Antarctic Bottom Water).

\begin{figure}
\centering
\includegraphics[width=0.8\textwidth]{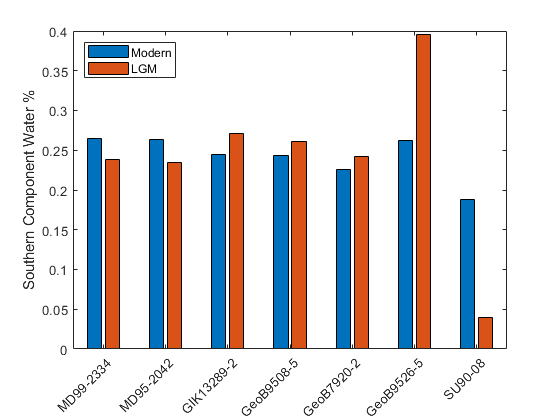}
\caption{Modern (blue) and LGM (red) Southern Sourced Water (SSW) percentages for the seven cores analyzed in the study.}
\label{figS-1-1}
\end{figure}

Figure \ref{figS-1-1} shows a present day model estimate \cite{Gebbie2012} and an LGM model estimate \cite{Oppo2018} of the fraction of SSW at these sites. The present day model suggests a similar composition across sites with slightly less SSW at SU90-08, which is farther north and west than the other sites. Modern-day ${\delta}^{18}{\rm O}$ values are very similar at these sites and are not greatly affected by the small difference in percent SSW. However, the water mass composition of these sites may have differed during the LGM. Currently available proxy data suggests that SU90-08 and GeoB9526-5 were bathed by 4\% and 40\% SSW respectively during the LGM. The water mass differences at these core sites could have produced inhomogeneous ${\delta}^{18}{\rm O}$ signals.

\begin{figure}
\centering
\includegraphics[width=0.8\textwidth]{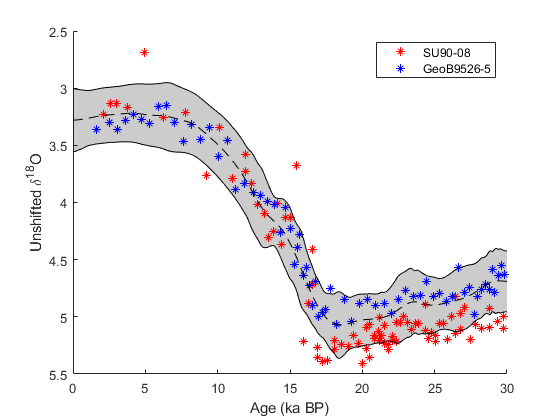}
\caption{Un-shifted ${\delta}^{18}{\rm O}$ records for SU90-08 (red) and GeoB9526-5 (blue) aligned to the DNEA stack (grey).}
\label{figS-1-2}
\end{figure}

Comparison of the un-shifted ${\delta}^{18}{\rm O}$ records from these cores provides another opportunity to assess whether they are homogeneous with the other cores in the DNEA stack (Figure \ref{figS-1-2}). While the ${\delta}^{18}{\rm O}$ record from GeoB9526-5 agrees well with the stack, the record from SU90-08 has higher ${\delta}^{18}{\rm O}$ values during the LGM. In addition, the dual proxy age for SU90-08 differs from the radiocarbon age (Figure 9 in the paper) during the deglaciation. In contrast, the two types of age models agree for GeoB9526-5.

Thus our initial assessment that SU90-08 is homogeneous with the DNEA stack remains inconclusive. The larger amplitude ${\delta}^{18}{\rm O}$ change in SU90-08 and the unusually large values at the start of the deglaciation are suggestive of a slightly different water mass history that may have resulted in an asynchronous ${\delta}^{18}{\rm O}$ signal. This may explain the slight discrepancy between dual proxy and radiocarbon ages (Figure 9 in the paper). When considering aligning two cores we suggest that researchers evaluate core locations with respect to water mass reconstructions and directly compare the features of the ${\delta}^{18}{\rm O}$ time series to evaluate whether the algorithm’s assumption of homogeneous ${\delta}^{18}{\rm O}$ variability is reasonable. In fact, analysis of many sites may provide a mechanism for evaluating different models of LGM water masses.

\section{Supplementary Notes: Details on SA-GPR}\label{S2}

Here we give details to the signal alignment and profile construction algorithms of the SA-GPR in section 2 of the main paper.

\subsection{Signal Alignment Algorithm}\label{S2-1}

For better understanding of the readers, in this section we simplify the notations used in the main body by dropping the superscript $\left(m\right)$, as follows:

\begin{itemize}
  \item ${\rm{X}}=\left\{x_n\right\}_{n=1}^{\rm{N}}$: inputs (or positions) of the signal.
  \item ${\rm{Y}}=\left\{y_n\right\}_{n=1}^{\rm{N}}$: outputs (or observations) of the signal.
  \item ${\rm{Z}}=\left\{{\rm{Z}}_n\right\}_{n=1}^{\rm{N}}$: hidden alignments of the signal to the profile.
  \item $\theta$: a set of parameters involved with the transition model.
  \item $\psi$: a set of parameters involved with the emission model.
  \item $\mathcal{P}$: the profile to be aligned.
\end{itemize}

Suppose that the transition and emission models (or the prior and likelihood) are the same as what are given in the main paper:
\begin{equation}
p\left({\rm{Z}}\middle|X;\theta\right)=\pi_0\left({\rm{Z}}_1\middle| x_1;\theta\right)\prod_{n=2}^{\rm{N}}{\pi\left({\rm{Z}}_n\middle|Z_{n-1},x_n,x_{n-1};\theta\right)},
\label{equS1}
\end{equation}
\begin{equation}
p\left({\rm{Y}}\middle|{\rm{Z}};\psi,\mathcal{P}\right)=\prod_{n=1}^{\rm{N}}{g\left(y_n\middle|\varphi\left({\rm{Z}}_n,\psi\right);\mathcal{P}\right)}.
\label{equS2}
\end{equation}

The alignment algorithm aims at sampling hidden alignments from its posterior distribution. The best case is to obtain the exact posterior of the hidden variable $p\left({\rm{Z}}\middle|{\rm{X}},{\rm{Y}};\theta,\psi,\mathcal{P}\right)\propto p\left({\rm{Z}}\middle|{\rm{X}};\theta\right)p\left({\rm{Y}}\middle|{\rm{Z}};\psi,\mathcal{P}\right)$ in a known closed form. However, it is in general impossible unless each ${\rm{Z}}_n$ takes at most finitely many values or both transition and emission models are Gaussian. The Markov-chain Monte Carlo algorithm is designed to sample from a posterior, even if it is not given in a closed form. However, the burn-in time is sensitive to the initialization of the hidden variable and crude proposal distributions often prevent efficient sampling. Therefore, the alignment algorithm of SA-GPR consists of two steps: the particle smoothing first initializes the hidden alignments and then the Metropolis-Hastings algorithm “refines” the initialized alignments.

The particle smoothing consists of two parts. The forward algorithm samples a set of candidates, or “particles”, from a proposal distribution $q_n$ at each step $n$, and computes weights on those particles to approximate the forward posterior with an empirical distribution. In formulation, it is expressed as follows:
\begin{equation}
p\left({\rm{Z}}_n\middle| x_{1:n},y_{1:n}\right) \approx \sum_{k=1}^{\rm{K}}{\omega_{n,k}1_{\left\{{\rm{Z}}_n=z_{n,k}\right\}}\left({\rm{Z}}_n\right)},
\label{equS3}
\end{equation}
where $\left\{z_{n,k}\right\}_{k=1}^{\rm{K}}$ are the sampled particles from $q_n$ at step $n$ and $\left\{\omega_{n,k}\right\}_{k=1}^{\rm{K}}$ are the associated weights sum to $1$.

Then, the forward posterior of the next step is updated iteratively as follows:
\begin{equation}
p \left({\rm{Z}}_{n+1}\middle| x_{1:n+1},y_{1:n+1}\right)\approx\sum_{k=1}^{\rm{K}}{\omega_{n+1,k}1_{\left\{{\rm{Z}}_{n+1}=z_{n+1,k}\right\}}\left({\rm{Z}}_{n+1}\right)},
\label{equS4}
\end{equation}
where $\left\{z_{n+1,k}\right\}_{k=1}^{\rm{K}} \sim_{i.i.d.} q_{n+1}$ and for $\sum_{k=1}^{\rm{K}}\omega_{n,k}=1$,
\begin{equation}
\omega_{n+1,k}\propto\frac{g\left(y_{n+1}\middle|\varphi\left(z_{n+1,k},\psi\right);\mathcal{P}\right)}{q_{n+1}\left(z_{n+1,k}\right)}\sum_{s=1}^{\rm{K}}{\omega_{n,s}\pi\left(z_{n+1,k}\middle| z_{n,s},x_{n+1},x_n;\theta\right)}.
\label{equS5}
\end{equation}

The backward algorithm samples each hidden alignment given the one after it as well as all inputs and outputs iteratively until a complete latent series is sampled. In formulation, it is expressed as follows:
\begin{equation}
p\left({\rm{Z}}_n=z_{n,k}\middle|{\rm{Z}}_n={\widetilde{z}}_{n+1},{\rm{X}},{\rm{Y}};\theta,\psi,\mathcal{P}\right)\propto\omega_{n,k}\cdot\pi\left({\widetilde{z}}_{n+1}\middle| z_{n,k},x_{n+1},x_n;\theta\right).
\label{equS6}
\end{equation}

Note that the particle smoothing is reduced to a hidden Markov model (HMM) if the proposal distribution is set to have the same finite support and all elements in the support are sampled once as particles. Also, because the particle smoothing does not compute the exact forward posterior, this method has limitations that HMMs do not. First, performance is dependent on the user-specified proposal distributions. Here, we iterate the sampling step consisting of the particle smoothing and Metropolis-Hastings and the parameter update step for $\theta$ and $\psi$ until convergence. Suppose that we obtained samples of the hidden alignments $\left\{{\widetilde{\rm{Z}}}^{\left(l\right)}\right\}_{l=1}^{\rm{L}}$ in the last round, where each ${\widetilde{\rm{Z}}}^{\left(l\right)}=\left\{{\widetilde{z}}_n^{\left(l\right)}\right\}_{n=1}^{\rm{N}}$. Then, each proposal $q_n$ at the current round is designed as follows:
\begin{equation}
q_n=\frac{1}{\rm{L}}\sum_{l=1}^{\rm{L}}1_{\left({\widetilde{z}}_n^{\left(l\right)}-d,{\widetilde{z}}_n^{\left(l\right)}+d\right)},
\label{equS7}
\end{equation}
where $d>0$ is a bandwidth hyperparameter and $\left({\widetilde{z}}_n^{\left(l\right)}-d,{\widetilde{z}}_n^{\left(l\right)}+d\right)$ is an interval. In other words, candidates at step n of the current round is randomly sampled from a randomly chosen interval $\left({\widetilde{z}}_n^{\left(l\right)}-d,{\widetilde{z}}_n^{\left(l\right)}+d\right)$ among $l=1,2,\cdots,{\rm{L}}$.

Second, a small number of output outliers might ruin the inference (especially if the transition model is too rigid). Lastly, the weights assigned to the particles are often too small to affect the backward sampling algorithm, which might cause a trouble in learning emission and transition models by the EM algorithm. These reasons prevent from relying only on the particle smoothing in sampling and just limit it as an initializer.

One advantage of the particle smoothing is that we can fast sample hidden alignments by the backward algorithm once particles and weights are obtained in the forward algorithm. To guarantee the independence, we first initialize the hidden alignments one-by-one by the particle smoothing and then run the Metropolis-Hastings algorithm on each of them.

The basic framework of the Metropolis-Hastings algorithm starts with computing the following acceptance ratio $\alpha$:
\begin{equation}
\alpha=\min{\left\{1,\frac{p\left({\rm{Y}}\middle|\dot{\rm{Z}};\psi,\mathcal{P}\right)p\left(\dot{\rm{Z}}\middle|{\rm{X}};\theta\right)}{p\left({\rm{Y}}\middle|{\rm{Z}};\psi,\mathcal{P}\right)p\left({\rm{Z}}\middle|{\rm{X}};\theta\right)}\cdot\frac{q\left({\rm{Z}}\middle|\dot{\rm{Z}}\right)}{q\left(\dot{\rm{Z}}\middle|{\rm{Z}}\right)}\right\}},
\label{equS8}
\end{equation}
where ${\rm{Z}}$ is the previously updated hidden alignment, $q\left(\cdot\middle|{\rm{Z}}\right)$ is the proposal distribution conditioned on ${\rm{Z}}$, and $\dot{\rm{Z}}$ is the proposed candidate that is sampled from $q\left(\cdot\middle|{\rm{Z}}\right)$. Then, update ${\rm{Z}}$ with $\dot{\rm{Z}}$ if $\alpha$ is bigger than or equal to a uniform random number in $\left[0,1\right]$; otherwise, keep ${\rm{Z}}$. Once we are in a burn-in phase, stop iteration and return the final ${\rm{Z}}$ as the sample.

Note that the Markov structure of transition model and conditionally independent emission model allow us to efficiently run the algorithm: propose simultaneously and decide whether the update for the odd $n$’s one-by-one and then do for the even $n$’s, and go back to the odd ones again.

To be more specific, the proposal distribution $q\left(\cdot\middle|{\rm{Z}}\right)$ is defined as follows:
\begin{equation}
q\left(\dot{\rm{Z}}\middle|{\rm{Z}}\right)=\prod_{n=1}^{\rm{N}}{q\left({\dot{\rm{Z}}}_n\middle|{\rm{Z}}\right)},
\label{equS9}
\end{equation}
\begin{equation}
q\left({\dot{\rm{Z}}}_n\middle|{\rm{Z}}\right) = \begin{cases}
        \mathcal{N}\left({\dot{\rm{Z}}}_n\middle|{\rm{Z}}_n,\frac{1}{8}\left({\rm{Z}}_{n+1}-{\rm{Z}}_n\right)\right),\ \ &n=1 \\ \mathcal{N}\left({\dot{\rm{Z}}}_n\middle|{\rm{Z}}_n,\frac{1}{8}\left({\rm{Z}}_n-{\rm{Z}}_{n-1}\right)\right),\ \ &n={\rm{N}} \\ \mathcal{U}\left({\dot{\rm{Z}}}_n\middle|{\rm{Z}}_{n-1},{\rm{Z}}_{n+1}\right),\ \ &otherwise \end{cases},
\label{equS10}
\end{equation}
where $\mathcal{U}\left({\dot{\rm{Z}}}_n\middle|{\rm{Z}}_{n-1},{\rm{Z}}_{n+1}\right)$ means that ${\dot{\rm{Z}}}_n$ follows a uniform distribution on the interval $\left({\rm{Z}}_{n-1},{\rm{Z}}_{n+1}\right)$.

\subsection{Profile Construction Algorithm}\label{S2-2}

This algorithm has been fully described in sections 2.1 and 2.2 of the main paper, so here we just discuss the heteroscedastic Gaussian process regression that we use in the simulations in section 2.3 and main results in section 4. The following method is from \cite{Lee2019}.

Suppose that we have obtained an explicit form of the Gaussian process regression for a given pair of sampled alignment and output $\left({\rm{Z}},{\rm{Y}}\right)$ as follows:
\begin{equation}
p\left(y\middle| z;{\rm{Z}},{\rm{Y}}\right)=\mathcal{N}\left(y\middle|\overline{\mu}\left(z\right),\overline{\nu}\left(z\right)+\Lambda\left(z\right)\right),
\label{equS11}
\end{equation}
where $\Lambda$ is the observational variance function. Note that $\overline{\mu}$ and $\overline{\nu}$ are defined once $\Lambda$ is given, as described in section 2.1.2. If $\Lambda$ is a constant function, we call the resulting model a homoscedastic Gaussian process regression; otherwise, it is a heteroscedastic Gaussian process regression. If $\Lambda$ is not given a priori and is supposed to be a nonconstant function, we define it based on the Nadaraya-Watson Kernel regression \cite{LANGRENE2019}:
\begin{equation}
\begin{aligned}
\Lambda\left(z\right) & \triangleq \mathbb{E}_{\left. f \middle| \mathcal{X}, \mathcal{Y} \right.} {\left[ \frac{\sum_{n=1}^{\rm N}{\left(y_n-{f}_{n} \right)^2 \mathcal{K}_h\left(x-x_n\right)}}{\sum_{n=1}^{\rm N}{\mathcal{K}_h\left(x-x_n\right)}} \right]}  \\ & = \frac{\sum_{n=1}^{\rm{N}}{\left(\left(y_n-\overline{\mu}\left({\rm{Z}}_n\right)\right)^2+\overline{\nu}\left({\rm{Z}}_n\right)\right)\mathcal{K}_h\left(z-{\rm{Z}}_n\right)}}{\sum_{n=1}^{\rm{N}}{\mathcal{K}_h\left(z-{\rm{Z}}_n\right)}},
\label{equS12}
\end{aligned}
\end{equation}
where $\mathcal{K}$ and $h>0$ are a density kernel and a bandwidth hyperparameter, respectively, and $f=\left\{f_n\right\}_{n=1}^{\rm{N}}$ is the realization of the latent continuous regression function assumed in the Gaussian process regression. Here $h$ can be tuned as a K-nearest neighborhood bandwidth \cite{terrell1992}.

Now the remaining issue is that $\Lambda$ should be given to compute $\overline{\mu}$ and $\overline{\nu}$ in (\ref{equS11}) while (\ref{equS12}) needs them to compute $\Lambda$. This can be resolved by iterating the following two steps until convergence: first update $\left(\overline{\mu}\left({\rm{Z}}_n\right),\overline{\nu}\left({\rm{Z}}_n\right)\right)$’s given $\Lambda\left({\rm{Z}}_n\right)$’s by (\ref{equS11}), and then update $\Lambda\left({\rm{Z}}_n\right)$’s given $\left(\overline{\mu}\left({\rm{Z}}_n\right),\overline{\nu}\left({\rm{Z}}_n\right)\right)$’s by (\ref{equS12}). Once all $\left(\overline{\mu}\left({\rm{Z}}_n\right),\overline{\nu}\left({\rm{Z}}_n\right)\right)$’s are trained together with $h$, then $\Lambda\left(z\right)$ can be defined for any $z$ by (\ref{equS12}).

\begin{figure}
\centering
\includegraphics[width=1\textwidth]{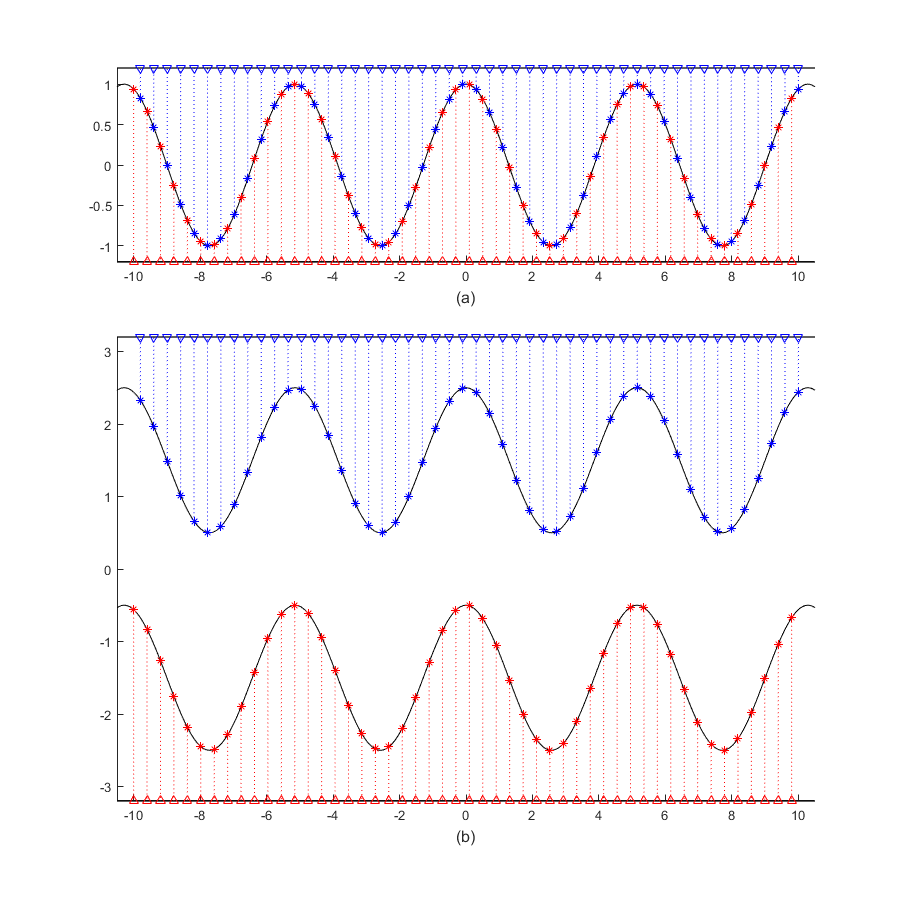}
\caption{(a) Two sets of input-output pairs, colored red and blue, are picked asynchronously from the same signal. (b) Two true signals to be aligned, which are designed to be identical.}
\label{fig2-1}
\end{figure}

\begin{figure}
\centering
\includegraphics[width=1\textwidth]{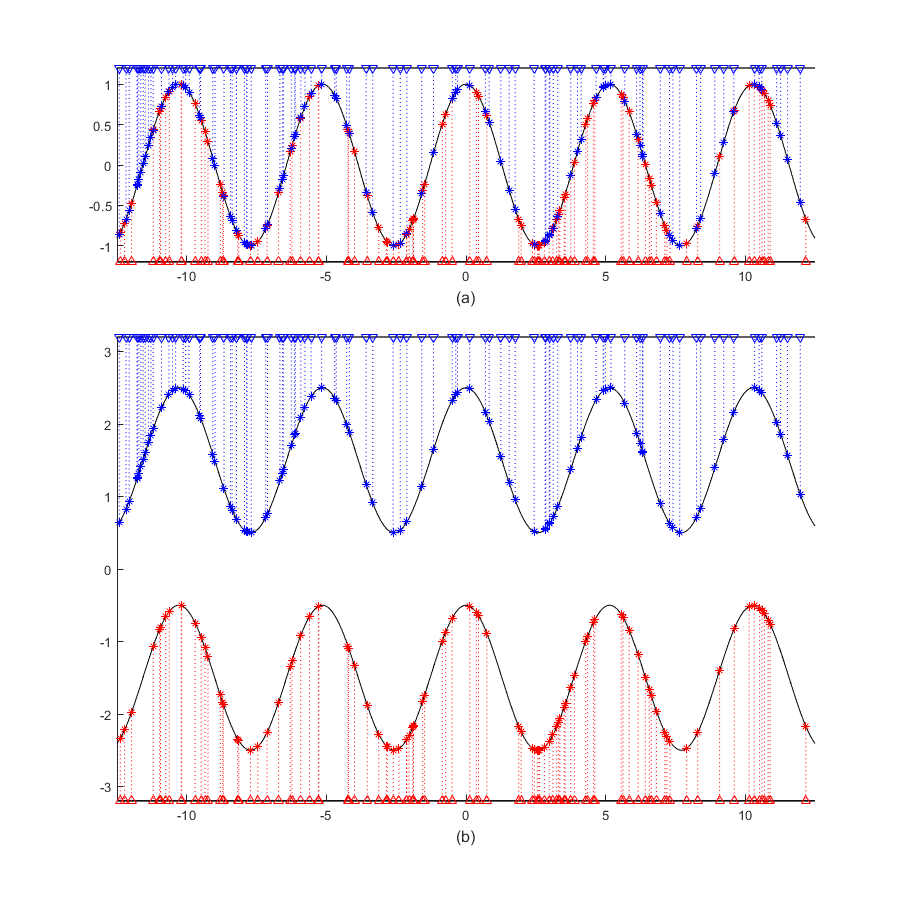}
\caption{(a) Two sets of input-output pairs, colored red and blue, are picked asynchronously from the same signal. (b) Two true signals to be aligned, which are designed to be identical.}
\label{fig2-3}
\end{figure}

\begin{figure}
\centering
\includegraphics[width=1\textwidth]{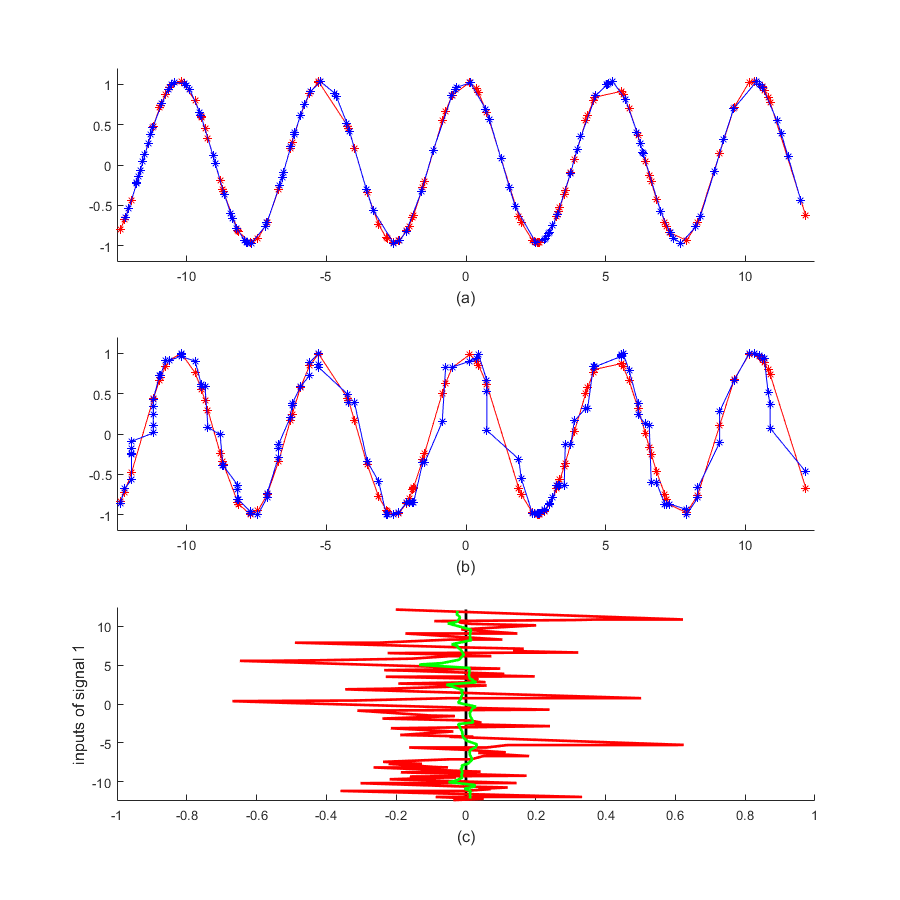}
\caption{(a) Median alignments obtained by SA-GPR. (b) Alignments obtained by DTW. (c) Errors of inferred alignments to the true ones. Red and green graphs are those of DTW and SA-GPR, respectively.}
\label{fig2-4}
\end{figure}

\begin{figure}
\centering
\includegraphics[width=0.8\textwidth]{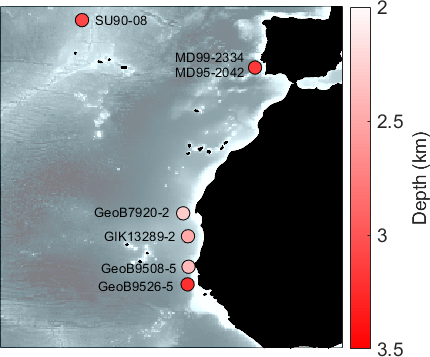}
\begin{tabular}{ccccc}
Core & Latitude & Longitude & Depth (m) & Citation \\
\hline
\hline \\ [-1.0em]
GeoB7920-2 & 20.75 & $-$18.58 & 2278 & \begin{tabular}{c} \cite{Tjallingii2008}, \\ \cite{Collins2011} \end{tabular} \\ [0.5ex]
\hline \\ [-1.0em]
GeoB9508-5 & 14.5 & $-$17.95 & 2384 & \cite{Mulitza2008} \\ [0.5ex]
\hline \\ [-1.0em]
GeoB9526-5 & 12.44 & $-$18.06 & 3223 & \begin{tabular}{c} \cite{Zarriess2011-1}, \\ \cite{Zarriess2011-2} \end{tabular} \\ [0.5ex]
\hline \\ [-1.0em]
GIK13289-2 & 18.07 & $-$18.01 & 2485 & \cite{Sarnthein1994} \\ [0.5ex]
\hline \\ [-1.0em]
MD95-2042 & 37.8 & $-$10.17 & 3146 & \begin{tabular}{c} \cite{Bard2004}, \\ \cite{Shackleton2000}, \\
\cite{Shackleton2004} \end{tabular} \\ [0.5ex]
\hline \\ [-1.0em]
MD99-2334 & 37.8 & $-$10.17 & 3146 & \begin{tabular}{c} \cite{Skinner2003}, \\ \cite{Skinner2004} \end{tabular} \\ [0.5ex]
\hline \\ [-1.0em]
ODP658C & 20.75 & $-$18.58 & 2273 & \cite{Demenocal2000} \\ [0.5ex]
\hline \\ [-1.0em]
SU90-08 & 43.35 & $-$30.41 & 3080 & \begin{tabular}{c} \cite{Paterne1999}, \\ \cite{Grousset1993} \end{tabular} \\ [0.5ex]
\end{tabular}
\caption{Locations of cores GeoB7920-2, GeoB9508-5, GeoB9526-5, MD95-2042, MD99-2334, GIK13289-2 and SU90-08.}
\label{fig4-1}
\end{figure}

\begin{figure}
\centering
\includegraphics[width=0.80\textwidth]{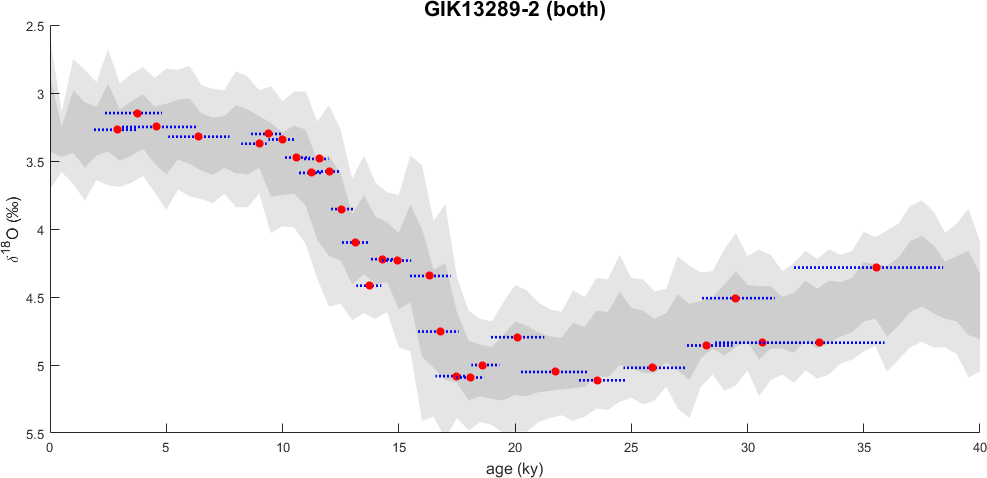}
\includegraphics[width=1.00\textwidth]{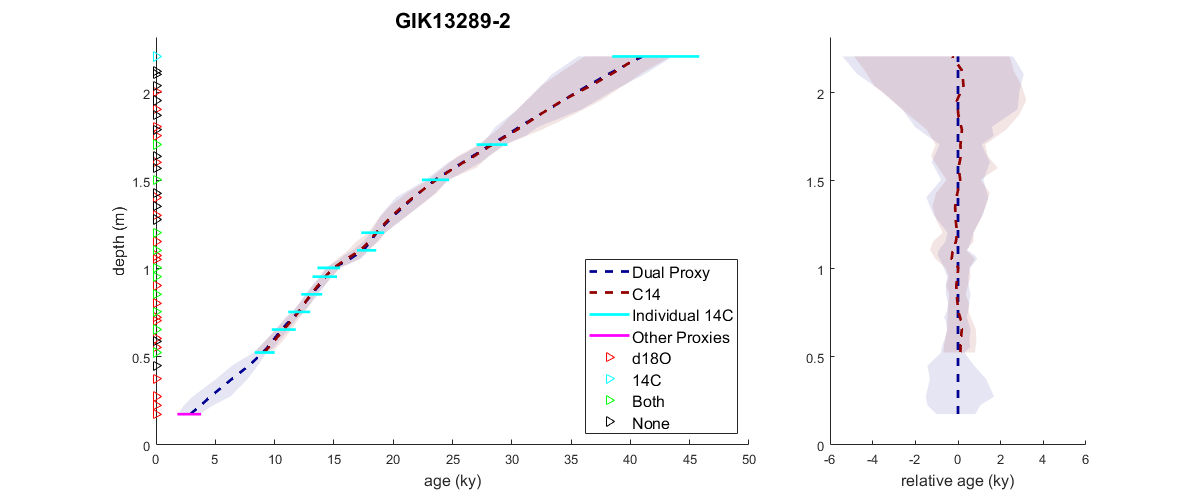}
\caption{Dual proxy age inferences and alignments of GIK13289-2 to DNA stack.}
\label{fig4-6}
\end{figure}

\begin{figure}
\centering
\includegraphics[width=0.80\textwidth]{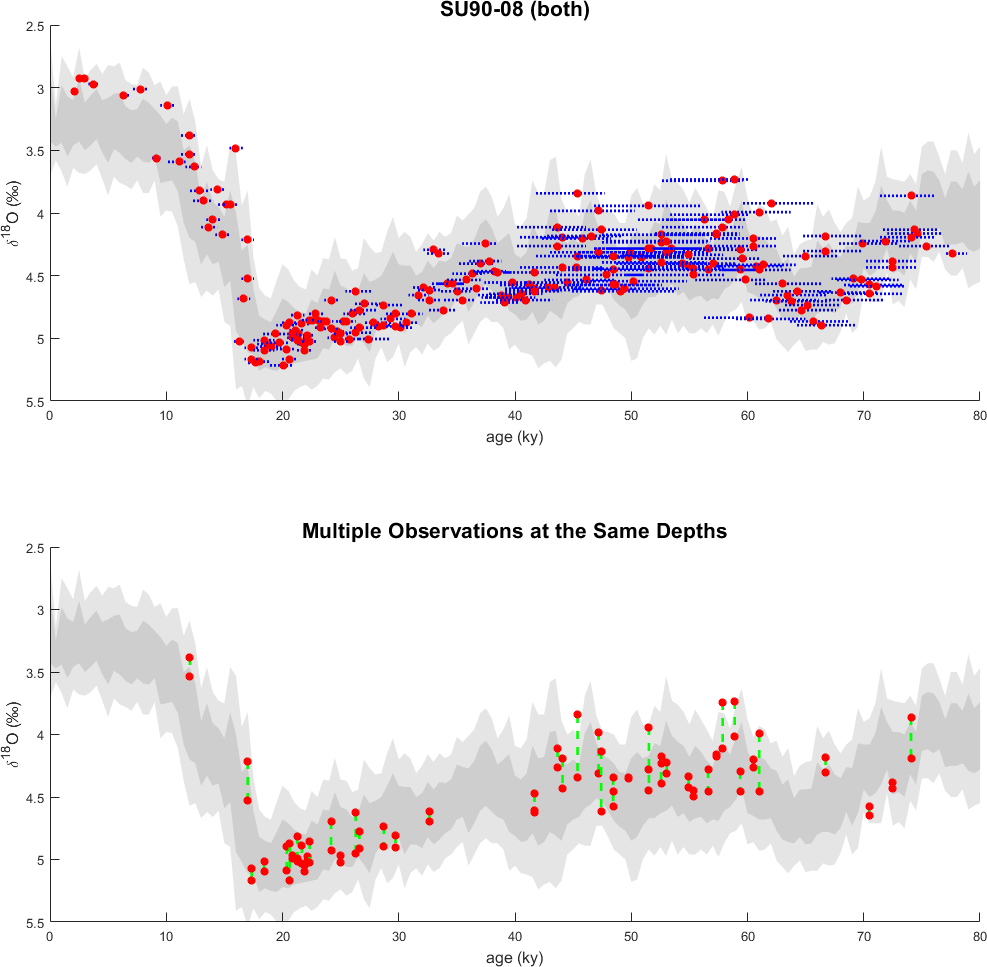}
\includegraphics[width=1.00\textwidth]{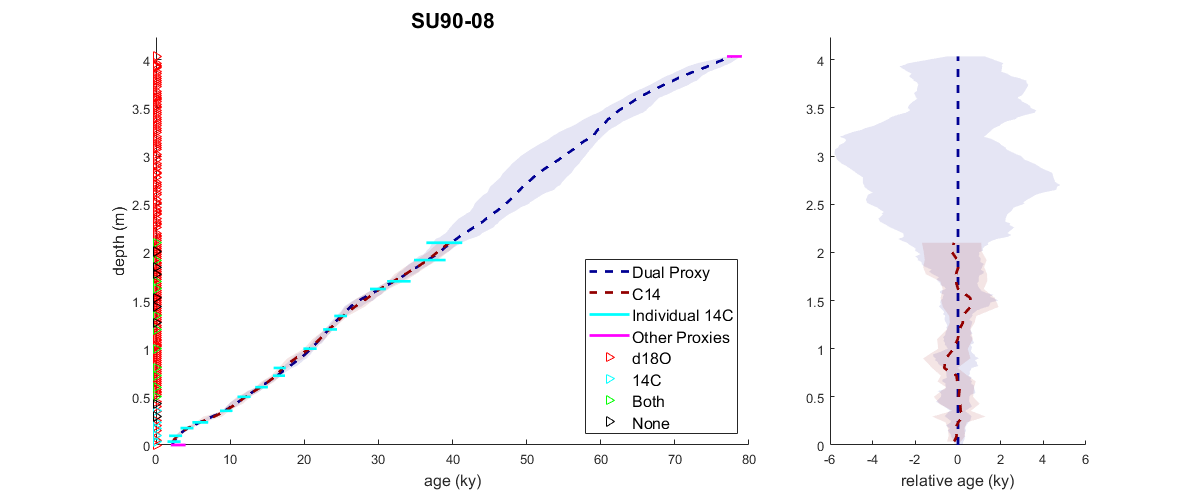}
\caption{Dual proxy age inferences and alignments of SU90-08 to DNA stack. This figure also shows how multiple observations at the same depths are dealt with in the alignment algorithm.}
\label{fig4-7}
\end{figure}

\section{Acknowledgements}\label{sec0}
This paper is based on the works supported by the National Science Foundation (NSF) under grant numbers OCE-1760838, OCE-1760878 and OCE-1760958, by the Division of Applied Mathematics in Brown University, and by the Kwanjeong Educational Foundation. Alan Jones assisted with data compilation.

\bibliographystyle{unsrt}
\bibliography{reference}

\end{document}